# Tunable Interlayer Charge-transfer States in $MoSe_2/WS_2$ Moiré Superlattices

Zheyu Lu[1,2,3,†,*], Jiahui Nie[1,2,3,4,†], Tianle Wang[1,†], Rwik Dutta[5,†], Ruishi Qi[1,3], Jingxu Xie[1,2,3,4], Can Uzundal[1,3,6], Jianghan Xiao[1,2,3], Ziyu Wang[1], Yibo Feng[1], Kenji Watanabe[7], Takashi Taniguchi[8], James R. Chelikowsky[5,9,10], Archana Raja[4,11], Steven G. Louie[1,3], Mit H. Naik[5], Michael P. Zaletel[1], Feng Wang[1,3,11,*]

[1]Department of Physics, University of California at Berkeley, Berkeley, California 94720, USA.
[2]Graduate Group in Applied Science and Technology, University of California at Berkeley, Berkeley, California 94720, USA.
[3]Material Science Division, Lawrence Berkeley National Laboratory, Berkeley, California 94720, USA.
[4]Molecular Foundry, Lawrence Berkeley National Laboratory, Berkeley, California 94720, USA.
[5]Department of Physics, The University of Texas at Austin, Austin, Texas 78712, USA.
[6]Department of Chemistry, University of California at Berkeley, Berkeley, California 94720, USA.
[7]Research Center for Electronic and Optical Materials, National Institute for Materials Science, 1-1 Namiki, Tsukuba 305-0044, Japan.
[8]Research Center for Materials Nanoarchitectonics, National Institute for Materials Science, 1-1 Namiki, Tsukuba 305-0044, Japan.
[9]Center for Computational Materials, Oden Institute for Computational Engineering and Sciences, The University of Texas at Austin, Austin, Texas 78712, USA.
[10]McKetta Department of Chemical Engineering, The University of Texas at Austin, Austin, Texas 78712, USA.
[11]Kavli Energy NanoScience Institute at University of California Berkeley and Lawrence Berkeley National Laboratory, Berkeley, California 94720, USA.

[†]These authors contributed equally.
[*]Correspondence to: zheyulu@berkeley.edu, fengwang76@berkeley.edu

# Abstract

Moiré superlattices formed by transition metal dichalcogenide (TMD) heterobilayers provide a versatile platform for studying strongly correlated electronic, excitonic, and topological phenomena in solids. In particular, angle-aligned $MoSe_2/WS_2$ heterobilayers, which have a Type-I band alignment at zero vertical electric field, host rich correlated spin and charge physics. Here, combining large-scale first-principles calculations and optical reflection spectroscopy, we report a thorough study of the emergent moiré excitonic states and interlayer charge-transfer states in angle-aligned electron-doped $MoSe_2/WS_2$ moiré superlattices. The moiré excitonic states serve as sensitive optical probes to the localization profile of doped electrons. We observe a series of interlayer charge-transfer transitions from $n/n_0 = 1$ to 4 (where $n_0$ denotes the moiré density) when the vertical electric field switches the heterostructure band alignment from Type-I to Type-II. By tuning the vertical electric field, we can precisely control the interlayer electron localization, realizing a Fermi-Hubbard model with a tunable charge-transfer band on an effective honeycomb lattice. Furthermore, Monte Carlo simulation of the doping dependence of the electric-field susceptibility predicts that multiple correlated charge-ordered states appear at both integer and fractional fillings. Our results provide a holistic understanding of the emergent optical excitations and the correlated charge-transfer states in electron-doped $MoSe_2/WS_2$ moiré superlattices.

# Main

The moiré superlattice, formed by stacking two-dimensional (2D) materials with a relative twist angle or lattice mismatch, offers a unique and powerful platform for studying strongly correlated electronic, excitonic, and topological physics in a highly tunable setting[1-44]. Specifically, $MoSe_2/WS_2$ moiré superlattices, with a Type-I band alignment without a vertical electric field, host fascinating correlated charge and spin physics[26-33]. Experimental study of charge physics, based on measuring the gate capacitance extracted from an additional optical sensing layer, has revealed a series of incompressible Wigner-Mott states[29]. Investigation of spin physics, based on measuring the spin susceptibility of the $n/n_0 = 1$ electron, has discovered signatures of kinetic magnetism near the electron-doped Mott insulator state[30]. However, the nature of multiple emergent optical transitions in the $MoSe_2/WS_2$ moiré superlattices in the electron-doped regime, as well as their strong electric field dependence, remained unclear. Here, we perform a thorough investigation of the doping- and electric field-dependence of the emergent moiré excitonic states in both 60°- (H-type) and 0°-aligned (R-type) $MoSe_2/WS_2$ moiré superlattices in the electron-doped regime. Using the emergent moiré excitonic states as optical probes, we demonstrate several robust electrically tunable electron-localization switching behaviors in the system at different filling factors. In electron-doped H-type $MoSe_2/WS_2$ moiré superlattices, charge transfer occurs between the $B^{Se/S}$ stacking in the $MoSe_2$ layer (M site) and the AB stacking in the $WS_2$ layer (W site), which forms a honeycomb lattice. This behavior is analogous to the interlayer charge transfer in hole-doped $WSe_2/MoTe_2$ heterostructures, which have been shown to host the unusual quantum anomalous Hall and Kondo physics[7,13]. Combined with Monte Carlo simulations, we further study the M-W site polarization susceptibility to the electric field in the $MoSe_2/WS_2$ moiré heterostructure and reveal a series of novel charge-ordered states.

Figure 1a shows a schematic cross-section and an illustration of the electrical circuit of our device D1. The $MoSe_2$ and $WS_2$ monolayers are stacked to be aligned at nearly 60° (H-type) and encapsulated by hexagonal boron nitride (hBN) on both top and bottom. Few-layer graphite flakes form the top electrostatic gate as well as the electrical contact, while 60 nm gold is used as the bottom electrostatic gate[33]. The dual-gate configuration enables independent control of the electron doping and the vertical electric field in the $MoSe_2/WS_2$ heterobilayer[33]. To achieve higher doping efficiency, we also apply an additional bottom-contact gate to heavily dope the device's electrical contacts, thereby reducing contact resistance at the moiré region.

The 4.4% lattice mismatch between $MoSe_2$ and $WS_2$ gives rise to a moiré superlattice with a period of $a_M \sim 8$ nm at a twist angle of 0° or 60°. Figure 1c shows the flat bands at the conduction band edge calculated using first-principles DFT calculations, with the gap corrected by GW calculations (see Supplementary Information). Figure 1b shows the modulation of the band state charge density arising from the lowest energy conduction band states derived from the K valley in $MoSe_2$ ($c_1$) and $WS_2$ ($c_7$) in the 60° $MoSe_2/WS_2$ moiré heterostructure. The emergence of flat electronic bands and modulation of the wavefunctions in real space leads to a modification of the optical absorption spectrum[32-34,43]. Figure 1d shows the reflection contrast (RC) spectrum of device D1 at charge neutrality with zero vertical electric field in the energy range around the $MoSe_2$ A excitons. The multiple absorption peaks are signatures of the large-area moiré superlattice[33]. We study the excitonic spectrum of the reconstructed moiré superlattice theoretically using GW-Bethe Salpeter equation calculations[45,46], which are made feasible in such a large superlattice using the pristine unit-cell matrix projection (PUMP) approach[33,34,43]. From our GW-BSE calculations using the PUMP method, the brightest peak ($M_1$ peak) is calculated to be strongly modulated in the moiré superlattice with the photoexcited hole and electron having the largest amplitude at the $B^{Se/S}$ stacking (which we are defining as the "M" site)[33], as shown in Figure 1e (see Supplementary Information for more details). Figure 1f and 1g display the RC spectra as a function of the electron doping (*y*-axis) at a vertical electric field at -0.07 V/nm and 0.07 V/nm, respectively. The electron doping is varied from $n/n_0$ = 0 to 4, where $n_0$ denotes the moiré density. Figure 1h (1i) shows the magnetic circular dichroism (MCD) spectra corresponding to Figure 1f (1g). Several new optical resonances are observed when the moiré heterostructure is electron-doped, and they exhibit nontrivial dependences on the electron density and the vertical electric field.

At low doping, a lower-energy peak first emerges (red arrow in Figure 1f). In 60°-aligned $MoSe_2/WS_2$ moiré superlattices, the first doped electron and the brightest intralayer moiré exciton of $MoSe_2$ ($M_1$ peak) are both localized at the M site (Figures 1b, e). As a result, a bound state composed of a doped electron and an exciton colocalized at the M site emerges at finite electron doping (Figure 1j). Therefore, we attribute this low-energy peak to the low-energy moiré trion (LET) state[33]. The significant positive MCD signal of the low-energy peak in Figures 1h and 1i, which is typical of a trion state, further validates this assignment. It is worth noting that there is another emergent excitation that is brighter than LET and has higher energy than the $M_1$ peak (Figures 1f and 1g). It exhibits the same evolutionary behavior as LET but lacks an intrinsic MCD signal (Figures 1h and 1i). We hypothesize that this high-energy peak arises from a higher-energy exciton state brightened by the presence of the doped electron. We refer to it as the high-energy moiré trion (HET) state (See Supplementary Information for more details). Interestingly, both LET and HET states show very different behaviors at -0.07 V/nm (Figure 1f) and 0.07 V/nm (Figure 1g).

Using the moiré trion (LET) as an optical probe, we systematically examine electron localization for doping ranging from $n/n_0$ = 0 to 3 at different electric fields. Figure 2a shows the two-dimensional map of the LET intensity, where the *x*-axis represents doping and the *y*-axis represents the vertical electric field (see Methods for a detailed convention about the direction of the vertical electric field). The LET resonance does not depend on the vertical electric field at $n/n_0 < 1$. However, its intensity switches at a threshold electric field for $1 < n/n_0 < 3$. To better illustrate this behavior, we show in Figure 2b two representative line cuts above and below the threshold from Figure 2a. In both cases, the intensity of the trion increases linearly from $n/n_0 = 0$ to 1 because the first electron occupies the $MoSe_2$ M site in the moiré supercell. As doping increases, more M sites are occupied by a single electron, resulting in a linear increase in trion intensity.

However, the evolution of behavior changes when we dope more than one electron per moiré supercell. When we apply an electric field below the threshold ($E$ = -0.07 V/nm), the trion intensity linearly decreases as the second electron also occupies the M site. The two electrons form a spin-singlet state, which prevents the formation of a moiré trion at the same position due to Pauli blocking. On the other hand, when we apply an electric field above the threshold ($E$ = 0.07 V/nm), the trion intensity does not change much between $n/n_0 = 1$ and 2. This observation indicates that the second electron does not stay at the $MoSe_2$ M site. Instead, the second electron occupies the AB site in $WS_2$ (W site). This charge-transfer process is a direct result of the competition between the conduction band offset of $MoSe_2$ and $WS_2$ (denoted by $\Delta$), which depends on the vertical electric field $E$ as $\Delta = \Delta_0 - edE$, and the onsite Coulomb repulsion of the M site in $MoSe_2$ (denoted by $U$). Here $\Delta_0$ is the zero electric field conduction band offset, $e$ is the electric charge, and $d$ is the interlayer distance. The switching happens at a critical electric field $E_c = (\Delta_0 - U)/(ed)$. In this device, $E_c$ = -0.003 V/nm. For charge densities between $n/n_0 = 2$ and 3, the trion intensity decreases linearly at $E$ = 0.07 V/nm. In this case, the third electron begins to fill the $c_1$ band and occupies the M site again after each W site in the moiré supercell has been filled by one electron.

At $E$ = -0.07 V/nm, a higher-energy peak emerges accompanying the disappearance of the LET resonance when $n/n_0 > 1$ (blue arrow in Figure 1f). Figure 2c shows the two-dimensional map of the intensity of this emergent excitation peak. This peak exists whenever two electrons occupy the M site, implying that this is the exciton state when the $c_1$ band is fully filled. In this sense, we call it the emergent moiré exciton (EX) state. Figure 2d shows two representative line cuts from Figure 2c at $E$ = -0.07 V/nm and 0.07 V/nm. At $E$ = -0.07 V/nm, the second doped electron occupies the M site and forms a spin-singlet state with the first doped electron between $n/n_0 = 1$ and 2. As a result, the trion intensity decreases linearly, whereas the EX intensity increases linearly, as shown by the blue curve in Figure 2d. Between $n/n_0 = 2$ and 3, the M site is filled with two electrons, and the third electron occupies the W site to avoid additional onsite Coulomb repulsion. As a result, the trion disappears, while EX persists. At $E$ = 0.07 V/nm, the second electron occupies the W site without affecting the trion intensity between $n/n_0 = 1$ and 2, and EX does not appear, as shown by the orange curve in Figure 2d. Between $n/n_0 = 2$ and 3, the third electron begins to occupy the M site, leading to a linear decrease in the trion intensity and a linear increase in the EX intensity. Therefore, EX serves as a complementary optical probe to identify electron localization in $MoSe_2/WS_2$ moiré superlattices, which can be precisely controlled by the vertical electric field.

In addition, at $n/n_0 < 1$, we can polarize the first doped electron to the W site at a high electric field. Figure 2g shows the two-dimensional map of the LET intensity in device D2, which has a higher breakdown electric field for the top and bottom gates and allows for a larger electric-field range in the measurement. Figure 2h shows the two-dimensional map of the $M_1$ intensity. At finite electron doping, when the electric field is larger than ~ 0.2 V/nm, LET disappears, and $M_1$ reappears, which indicates that the doped electron transfers to the W site, leaving the M site vacant. Such switching occurs at another critical electric field $E_c' = \Delta_0/(ed)$ and represents the Type-I to Type-II transition in $MoSe_2/WS_2$ moiré superlattices.

Using LET and EX as optical probes, we established several electrically tunable electron-localization switching behaviors in the $MoSe_2/WS_2$ moiré superlattices at different doping regimes (See additional switching behaviors in Supplementary Information): (1) $n/n_0 < 1$: The electron localization can be switched between the $B^{Se/S}$ stacking (Figure 1b) in the $MoSe_2$ layer (M site) and the AB stacking (Figure 1b) in the $WS_2$ layer (W site) by applying a large positive electric field to compensate $\Delta_0$. A honeycomb lattice is formed when we tune the electric field to make the two sites energetically degenerate. (2) $1 < n/n_0 < 3$: The first electron occupies the M site and forms a triangular lattice, while the second electron can be switched between the M and W sites by applying a moderate vertical electric field. In this case, a honeycomb lattice for the second electron can also be formed.

A similar charge-transfer behavior is also observed in 0°-aligned (R-type) $MoSe_2/WS_2$ moiré superlattices, as shown in Figure 3. However, the critical electric field for the charge-transfer transition is different. Figure 3a and 3b show the lowest flat bands at the conduction band edge and the associated charge density distribution in the R-type $MoSe_2/WS_2$ moiré superlattices obtained from our first-principles calculations. The lowest conduction band derived from the K valley in the $MoSe_2$ layer is the $c_1$ band with electrons localized at the AA stacking site. The lowest conduction band derived from the K valley in the $WS_2$ layer is the $c_{10}$ band with electrons localized at the $B^{Se/W}$ stacking site.

Figure 3c shows the doping-field RC spectrum of the $MoSe_2$ layer in device D3 with an R-type $MoSe_2/WS_2$ moiré superlattice. The density dependence of the $MoSe_2$ RC spectra exhibits strong changes as the electric field increases from 0.22 V/nm to 0.4 V/nm. At all vertical electric fields, the $MoSe_2$ LET signal (denoted by the blue arrow) increases from charge neutral as the electron doping increases and reaches a maximum at $n/n_0 = 1$. This indicates that the first doped electron is localized at the AA site of the $MoSe_2$ layer, suggesting a Type-I band alignment in the applied electric field range. At a low field of 0.22 V/nm, the $MoSe_2$ LET signal starts to decrease when $n/n_0 > 1$ and disappears at $n/n_0 = 2$. It suggests that the second doped electron is also localized at the same site in the $MoSe_2$ layer. The two electrons form a singlet pair and suppress the $MoSe_2$ LET signal. When the electric field is increased to 0.263 V/nm, the LET signal remains constant between $1 < n/n_0 < 2$ and starts to decrease after $n/n_0 > 2$. It suggests that the second electron is now transferred to the $WS_2$ layer, sitting at the $B^{Se/W}$ site, while the third electron is doped in the $MoSe_2$ layer, occupying the AA site again. At a higher electric field of 0.360 V/nm, the LET signal remains constant between $1 < n/n_0 < 3$ and starts to decrease after $n/n_0 > 3$. It indicates that both the second and the third electrons are transferred to the $WS_2$ layer and the fourth electron occupies the $MoSe_2$ layer again. At even higher electric field of 0.40 V/nm, the second, third, and fourth electrons are all in the $WS_2$ layer, and only the fifth electron occupies the $MoSe_2$ layer again, as shown by the constant LET signal between $1 < n/n_0 < 4$. These unusual charge-transfer behaviors are also manifested in the corresponding changes of the $MoSe_2$ EX signals (denoted by the green arrow).

Figure 3d shows the peak intensity of $MoSe_2$ LET (top) and EX (bottom) states across the full doping and electric field range. We can clearly see the charge-transfer transitions of $n = 2,3,4$, which take place at a threshold electric field of around 0.272 V/nm, 0.298 V/nm, and 0.370 V/nm, respectively.

Next, we quantitatively analyze the electric field induced charge-transfer transition in device D1, as reflected in the changes in the LET and EX intensities over the doping range $1 \leq n/n_0 \leq 3$ from Figure 2. At each doping, the intensity evolution of both excitations is well described by a polarization function (See Supplementary Information for more details):

$$I = f \cdot \tanh(\alpha \cdot (E - E_c)) + c$$

where $c$ and $f$ set the intensity offset and contrast, $\alpha$ describes its field susceptibility, and $E_c$ is the critical electric field at its midpoint. Figure 4a presents the doping dependence of the four fitting parameters for both LET (blue curves) and EX (orange curves). The transition behaviors extracted from these two optical probes are in good agreement.

To understand the doping dependence of the fitting parameters, we first consider a minimal two-site model with M and W sites in each moiré supercell. In the absence of interactions, the moiré trion intensity at $1 < n/n_0 < 2$ is given by

$$I \propto (1 - \delta/2) + \delta/2 \cdot \tanh(\beta(ehE - \Delta)/2)$$

where $\delta = n/n_0 - 1$ is the doping measured from $n/n_0 = 1$, $\beta = 1/(k_B T)$ is the thermodynamic beta, $h$ is the effective M-W vertical separation that determines their energy difference at electric field $E$, and $\Delta$ is the free energy difference for adding one more electron on M relative to W (See Supplementary Information for details). A similar relation applies to $2 < n/n_0 < 3$:

$$I \propto (1 - \delta/2) + (1 - \delta/2) \cdot \tanh(\beta(ehE - \Delta)/2).$$

This simple model naturally explains the linear dependence of $f$ and $c$ on doping. The intensity contrast $f$ measures the number of carriers that can be polarized between the two sites. It therefore increases linearly for $1 < n/n_0 < 2$, reaches a maximum at $n/n_0 = 2$, and then decreases linearly for $2 < n/n_0 < 3$ as more moiré supercells enter the less-polarizable three-electron configuration. Meanwhile, the intensity offset $c$ corresponds to the half-height intensity at $E = E_c$, where the M and W sites are degenerate, and the added electron occupies them with equal probability. The linear doping dependence of $c$ reflects the growing fraction of moiré supercells with two electrons on the M site, which decreases (increases) the LET (EX) intensity.

However, the non-interacting two-site model predicts a doping-independent susceptibility, inconsistent with the experimental trend. This discrepancy indicates that Coulomb interactions induce correlated polarization across the moiré lattice. To capture this effect, we include long-range Coulomb interactions in our model and use Monte Carlo simulations to compute the doping-dependent electron configurations and the corresponding M-W polarization susceptibility $\alpha$. The interacting model is:

$$H(n) = \frac{1}{2}\sum_{ij} V_{ij} n_i n_j - gE_z \sum_i z_i n_i$$

According to the DFT calculation of the electron-localization profile (Figure 1b), we place the M and W sites on the two sublattices of an effective honeycomb lattice, with $z_i = +1$ (-1) for M (W) sites. $V_{ij}$ is the screened Coulomb potential between site $i$ and $j$. Each site has occupation $n_i \in \{0, 1\}$, defined relative to the $n/n_0 = 1$ background charge with one electron on each M site, so that the model is particle-hole symmetric about $n/n_0 = 2$ (below, we focus on $1 < n/n_0 < 2$). Because this subtraction introduces an entropic contribution that shifts the free energy of each site, we include the corresponding term explicitly in the Monte Carlo sampling weights to recover the correct physical energetics. For simplicity, we neglect coherent electron hopping and treat the model as classical; this approximation is justified because the kinetic energy is much smaller than the Coulomb scale and does not qualitatively change the charge configurations relevant to the polarization response. We also neglect the layer dependence of the Coulomb interaction between M-M, M-W, and W-W sites, which results in a doping-independent $E_c$ in our simulations. In contrast, the doping dependence and particle-hole asymmetry of $E_c$ in the experiment can be attributed to differences in the site/layer dielectric environment between the M and W sites.

We extract $\alpha$ from our simulation over a wide range of temperatures (Figure 4b). At moderately low temperatures, the simulation $\alpha(n)$ captures the nonmonotonic trend observed in our experiment. Near $n/n_0 = 1$, $\alpha$ approaches the noninteracting two-level susceptibility $\alpha_{\mathrm{T}} = \beta e h/2$, but it decreases gradually as $n/n_0$ increases, indicating suppression of collective polarization from repulsive interactions. After reaching a minimum near $n/n_0 = 3/2$, $\alpha$ rises sharply and diverges as $n/n_0 \rightarrow 2$, signaling the emergence of a polarized charge-ordered state at $n/n_0 = 2$ stabilized by Coulomb interaction. The resulting state is analogous to the correlated insulating phases discussed previously in triangular-lattice moiré systems at fractional fillings[3]. We also examine the temperature evolution of $\alpha$: At lower temperatures, additional sharp non-monotonic features appear in $\alpha(n)$, suggestive of a "devil's staircase" of competing commensurate charge orders. At higher temperatures, the strong enhancement of $\alpha$ at $n/n_0 = 2$ is rounded, consistent with thermal melting of the charge-ordered state. Moreover, for all dopings away from $n/n_0 = 1$, $\alpha(T)$ is generally nonmonotonic and deviates from the noninteracting expectation. This behavior confirms the essential role of interactions in explaining the measured susceptibility.

We further analyze the zero-field electron configurations sampled in the simulations to identify the associated charge orders. As temperatures are lowered, we observe an increasing number of commensurate charge patterns, whose Monte Carlo trajectories exhibit markedly longer autocorrelation times (Figure 4c), indicating slow dynamics and incipient charge ordering. The $n/n_0 = 2$ charge order appears first and corresponds to half-filling of the effective honeycomb lattice. In this state, electrons are fully polarized onto a single sublattice, which explains its exceptionally strong response to the applied polarizing field. The next robust charge order occurs at $n/n_0 = 4/3$, the 1/6 filling of the effective honeycomb lattice, and exhibits a similarly charge-polarized state with a tripled supercell. The enhancement of susceptibility remains unresolved in our experiment, possibly because the ordering temperature is lower. Finally, near $n/n_0 = 3/2$, we identify a prominent local charge pattern. This pattern is stripe-like, with electrons alternating between M and W sites, thereby breaking the underlying $C_3$ rotational symmetry. We do not observe unambiguous long-range stripe order within the accessible temperature window. Still, it can nonetheless account for the suppression of $\alpha$ near $n/n_0 = 3/2$: alternating M/W occupation penalizes field-driven relocation of individual electrons and reduces the collective polarization response.

In summary, we have reported optical measurements of emergent moiré excitations in the doping and electric field two-dimensional phase space for both electron-doped H-type and R-type $MoSe_2/WS_2$ moiré superlattices. Our results provide a comprehensive understanding of the nature of these excitations, which previously remained unclear. The moiré trion and the emergent moiré exciton serve as optical probes to identify electron localization and interlayer charge transfer, which can be precisely controlled by a vertical electric field. These novel electron-localization switching behaviors, along with tunable lattice geometries between triangular and honeycomb configurations, enable the realization of a Fermi-Hubbard model with an electric field-tunable charge-transfer band in $MoSe_2/WS_2$ moiré superlattices. Furthermore, $MoSe_2/WS_2$ moiré superlattices have the potential to host ferroelectric and Wigner-molecule states. Our work paves the way for future experiments that utilize moiré excitations as optical probes to explore electron localization in TMD moiré superlattices.

# Methods

**Device fabrication**
We use a dry-transfer method based on polyethylene terephthalate glycol (PETG) stamps to fabricate the heterostructures. Monolayer $MoSe_2$, monolayer $WS_2$, few-layer graphene, and hBN flakes are mechanically exfoliated from bulk crystals onto Si substrates with a 90-nm-thick $SiO_2$ layer. We use 20-30 nm hBN as the gate dielectric. A 0.5 mm-thick clear PETG stamp is employed to pick up the flakes at 65-75°C sequentially. The whole stack is then released onto a 285 nm $SiO_2$/Si substrate with a pre-patterned 60 nm Au bottom gate (D1 and D2) and a 10 nm Pt bottom gate plus a contact gate (D3) at 95-100°C, followed by dissolving the PETG in chloroform at room temperature for at least one day. Electrodes (Cr/Au for D1 and D2; Bi/Sb/Au for D3) are defined using photolithography (Durham Magneto Optics, MicroWriter) or electron-beam lithography (Crestec) and evaporated using electron-beam evaporation (Angstrom Engineering).

**Doping- and electric field-dependent RC spectroscopy**
The optical measurements are performed in an optical cryostat (Quantum Design, OptiCool) with a temperature down to 2 K (nominal). We use a supercontinuum laser (Fianium Femtopower 1060 Supercontinuum Laser) as the light source for the reflection spectroscopy. The laser is focused on the sample using a 20× Mitutoyo objective with a beam diameter of ~ 1.5 µm. A small beam size provides a local probe that we can park in a clean region free of bubbles. We use a very low incident laser power (approximately 10 nW) to minimize photodoping effects. The spectra are independent of the incident light power up to 200 nW. The reflected light is collected by a spectrometer (Princeton Instruments PIXIS 256e) with a 1000 ms exposure time. To minimize the influence of laser fluctuations, a second laser beam, directly reflected from a silver mirror, is collected simultaneously to normalize the sample reflection spectra.

Doping- and electric field-dependent measurements are conducted by applying voltages to the graphite and gold gates using Keithley 2400 SourceMeters. The charge density $n$ and the vertical electric field $E$ are defined using a parallel plate capacitor model:

$$n = \frac{1}{e}\frac{\epsilon_{\mathrm{hBN}}\epsilon_0}{d}(V_\mathrm{T} + V_\mathrm{B} - V_0)$$

$$E = \frac{1}{2d}(V_\mathrm{T} - V_\mathrm{B})$$

where $\epsilon_{\mathrm{hBN}}$ is the dielectric constant of hBN, $\epsilon_0$ is the permittivity of free space, $d$ is the thickness of the top and bottom hBN layer (we choose two hBNs with the same thickness to make the device symmetric to both gates), and $V_\mathrm{T}$ ($V_\mathrm{B}$) is the voltage applied to the top (bottom) gate. We account for the quantum capacitance (the voltage range over which charge is not injected) by applying an offset voltage $V_0$ for electron doping. The offsets are defined by the sum voltage at which the reflection contrast spectrum begins to change with electron doping. The vertical electric field uses a sign convention in which a positive vertical electric field points from the top gate to the bottom gate.

The moiré density $n_0$ is defined to correspond to one electron per moiré supercell and is determined through the relation $n_0 = 1/[L_\mathrm{M}^2 \sin(\pi/3)]$ where $L_\mathrm{M}$ is the moiré superlattice constant. The twist angle ($\theta$) and lattice mismatch ($\delta = (a' - a)/a$) between the two layers determine $L_\mathrm{M}$ via $L_\mathrm{M} = a/\sqrt{\delta^2 + \theta^2}$.

**Magnetic circular dichroism (MCD) measurements**

The MCD data, defined as $(R_+ - R_-)/(R_+ + R_-)$, were taken at $B$ = 0.5 T. Here $R_{+/-}$ denotes the reflectivity of two circular polarizations.

**Numerical simulation**

We perform classical Monte-Carlo simulations of our two-site model on the effective honeycomb lattice. The simulation is performed on up to 30 × 30 unit cells, and numerical convergence with respect to system size is confirmed. For $V_{ij}$ we use the double-gate-screened Coulomb interaction with $d/2$ the gate-to-sample distance: $V(r_{ij}) = \frac{e^2}{4\pi\epsilon_r\epsilon_0}\sum_{k=-\infty}^{+\infty}\frac{(-1)^k}{\sqrt{r^2+(kd)^2}}$. The interaction decays as $r^{-3}$ over $r$, and we truncate terms smaller than 5% of the nearest-neighbor interaction $V_1$. Based on our experiment, we choose the moiré unit length $a_M$ = 8 nm, gate distance $d$ = 60 nm, and dielectric constant $\epsilon_r$ = 8, which gives $V_1 = 34$ meV $= k_B \cdot 400$ K.

**First-principles calculations**

We investigate the structural reconstruction, electronic structure, and optical absorption of $MoSe_2/WS_2$ moiré superlattices at twist angles of 0° and 60° using first-principles calculations. The atomic structures of these superlattices are first relaxed using classical interatomic force-field methods[47,48,49]. To analyze intralayer moiré excitonic absorption associated with the $MoSe_2$ layer of the superlattice, we compute the electronic structure of the reconstructed $MoSe_2$ layer within density functional theory (DFT). Excitonic properties and optical absorption spectra are then obtained using many-body perturbation theory within the GW approximation combined with the Bethe–Salpeter equation (GW-BSE). The BSE Hamiltonian is constructed using 24 valence and 24 conduction bands, sampled on a 3 × 3 × 1 $k$-point grid of the moiré Brillouin zone. The moiré BSE kernel is evaluated using the pristine unit-cell matrix projection (PUMP) approach, where the moiré kernel matrix elements are expressed as a coherent linear combination of pristine unit-cell kernel matrix elements (See Supplementary Information for more details).

# Data availability

The main data that support the findings of this study are available within the article and its Supplementary Information files. More supporting data are available from the corresponding authors upon request.

**Acknowledgements**

This work was supported primarily by the Center for Computational Study of Excited-State Phenomena in Energy Materials (C2SEPEM) at LBNL, funded by the U.S. Department of Energy, Office of Science, Basic Energy Sciences, Materials Sciences and Engineering Division, under Contract DE-AC02-05CH11231, as part of the Computational Materials Sciences Program. The heterostructure device fabrication is supported by the Office of Basic Energy Sciences, Materials Sciences and Engineering Division, of the U.S. Department of Energy under Contract DE-AC02-05-CH11231, within the van der Waals Heterostructures Program (KCWF16). This research used the Lawrencium computational cluster resource provided by the IT Division at the Lawrence Berkeley National Laboratory (Supported by the Director, Office of Science, Office of Basic Energy Sciences, of the U.S. Department of Energy under Contract No. DE-AC02-05CH11231). M.H.N. acknowledges support from the National Science Foundation (NSF) MRSEC DMR-2308817. R.D. was supported by the Texas Quantum Institute Graduate Fellowship. K.W. and T.T. acknowledge support from the JSPS KAKENHI (Grants 21H05233 and 23H02052), CREST (JPMJCR24A5), JST, and World Premier International Research Center Initiative, MEXT, Japan. We acknowledge the Texas Advanced Computing Center at The University of Texas at Austin for providing computational resources that have contributed to the research results reported within this paper. This work also used computational resources from Stampede3 at The University of Texas at Austin through allocation PHY250206 from the Advanced Cyberinfrastructure

Coordination Ecosystem: Services and Support program, which is supported by NSF Grants 2138259, 2138286, 2138307, 2137603, and 2138296.

**Author contributions** Z.L. and F.W. conceived the research. Z.L. and R.Q. carried out optical measurements with J.N., J. Xie, C.U., and A. Raja. Z.L., J.N., and F.W. performed data analysis. T.W., Z.L., and M.Z. performed Monte Carlo simulations. R.D., J.R.C., S.G.L., and M.H.N. performed first-principles DFT and GW-BSE calculations. Z.L., J.N., J. Xie, J. Xiao, Z.W., and Y.F. contributed to the fabrication of the van der Waals heterostructures. K.W. and T.T. grew hBN crystals. All authors discussed the results and wrote the manuscript.

**Competing interests** The authors declare no competing interests.

# Figures

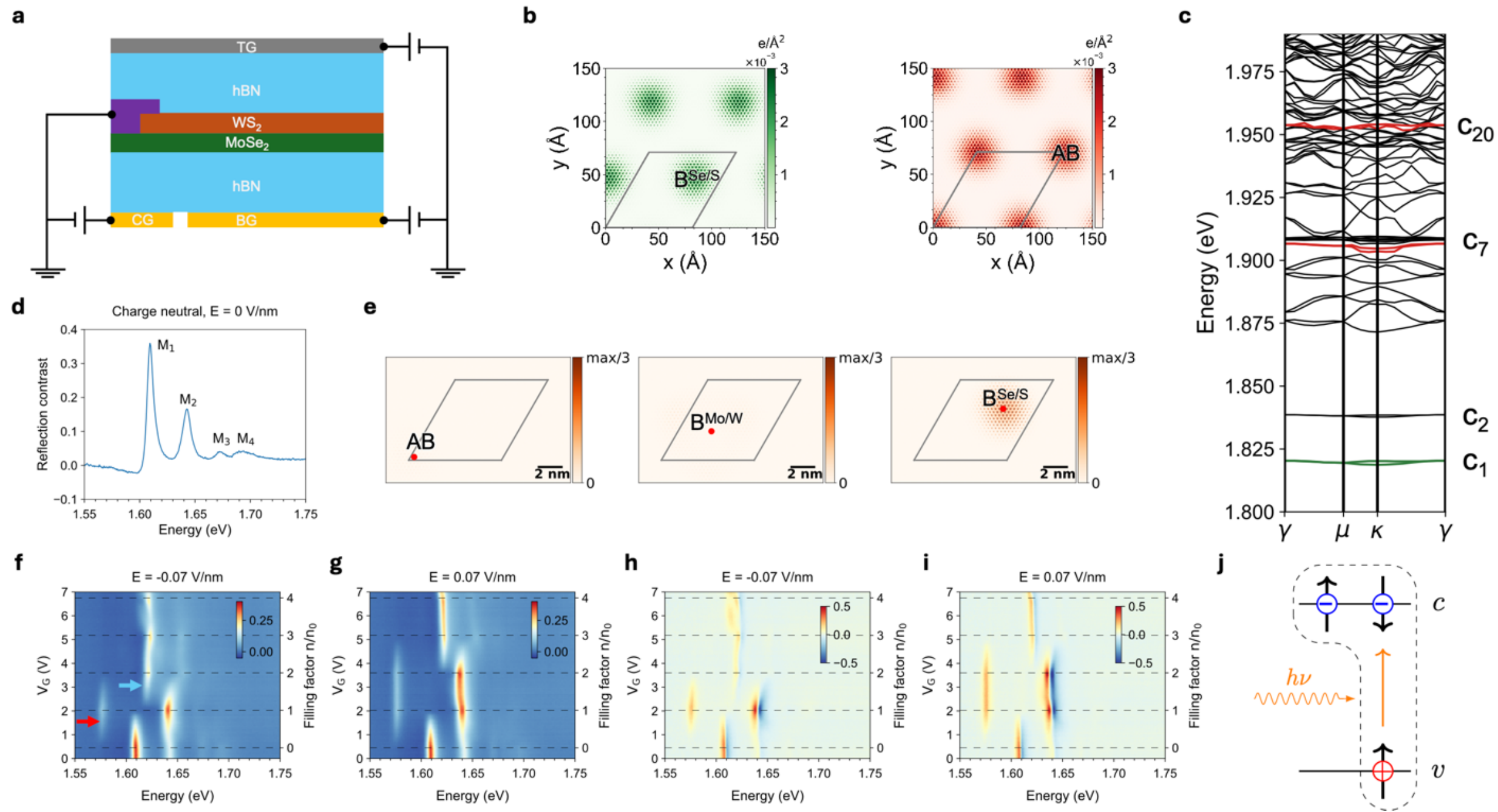

**Fig. 1 | Emergent moiré excitations in H-type $MoSe_2/WS_2$ moiré superlattices. (a)** Schematic showing the device structure. **(b)** First-principles calculations of the electron charge density modulation in H-type $MoSe_2/WS_2$ moiré superlattices, taking into account structural reconstructions in the bilayer. Left: charge density of the lowest moiré conduction band state derived from the K valley of the $MoSe_2$ layer (marked as $c_1$ band) is localized at the $B^{Se/S}$ stacking; right: charge density of the lowest energy conduction band derived from the K valley of the $WS_2$ layer (marked as $c_7$ band). **(c)** Electronic band structure of the reconstructed H-type $MoSe_2/WS_2$ moiré superlattices calculated using DFT, with the band gap corrected using GW calculations. $c_1$ and $c_2$ bands are localized in the $MoSe_2$ layer; $c_7$ and $c_{20}$ bands are the first two lowest energy conduction states localized in the $WS_2$ layer, derived from the K valley. **(d)** Experimental reflection contrast (RC) spectrum of H-type $MoSe_2/WS_2$ moiré superlattices under charge neutrality and zero external vertical electric field. The four main moiré exciton states are labeled as $M_1$, $M_2$, $M_3$, and $M_4$. **(e)** Calculated real-space map of the moiré exciton corresponding to the brightest $M_1$ exciton state in our first-principles GW-BSE calculations. The plot shows the electron density of the exciton, $|\chi(\mathbf{r_e}, \mathbf{r_h})|^2$, for fixed hole positions $\mathbf{r_h}$ (red dot) in the superlattice. **(f-g)** Experimental RC spectrum of H-type $MoSe_2/WS_2$ moiré superlattices with electron doping ranging from $n/n_0 = 0$ to 4 under a moderate negative (f) and a positive (g) electric field, respectively. $V_G = (V_T + V_B) / 2$. The red and blue arrows indicate the low-energy moiré trion (LET) and emergent moiré exciton (EX) states, respectively. **(h-i)** Experimental magnetic circular dichroism (MCD) spectrum corresponding to (f-g). **(j)** Schematic showing the physical nature of the LET state.

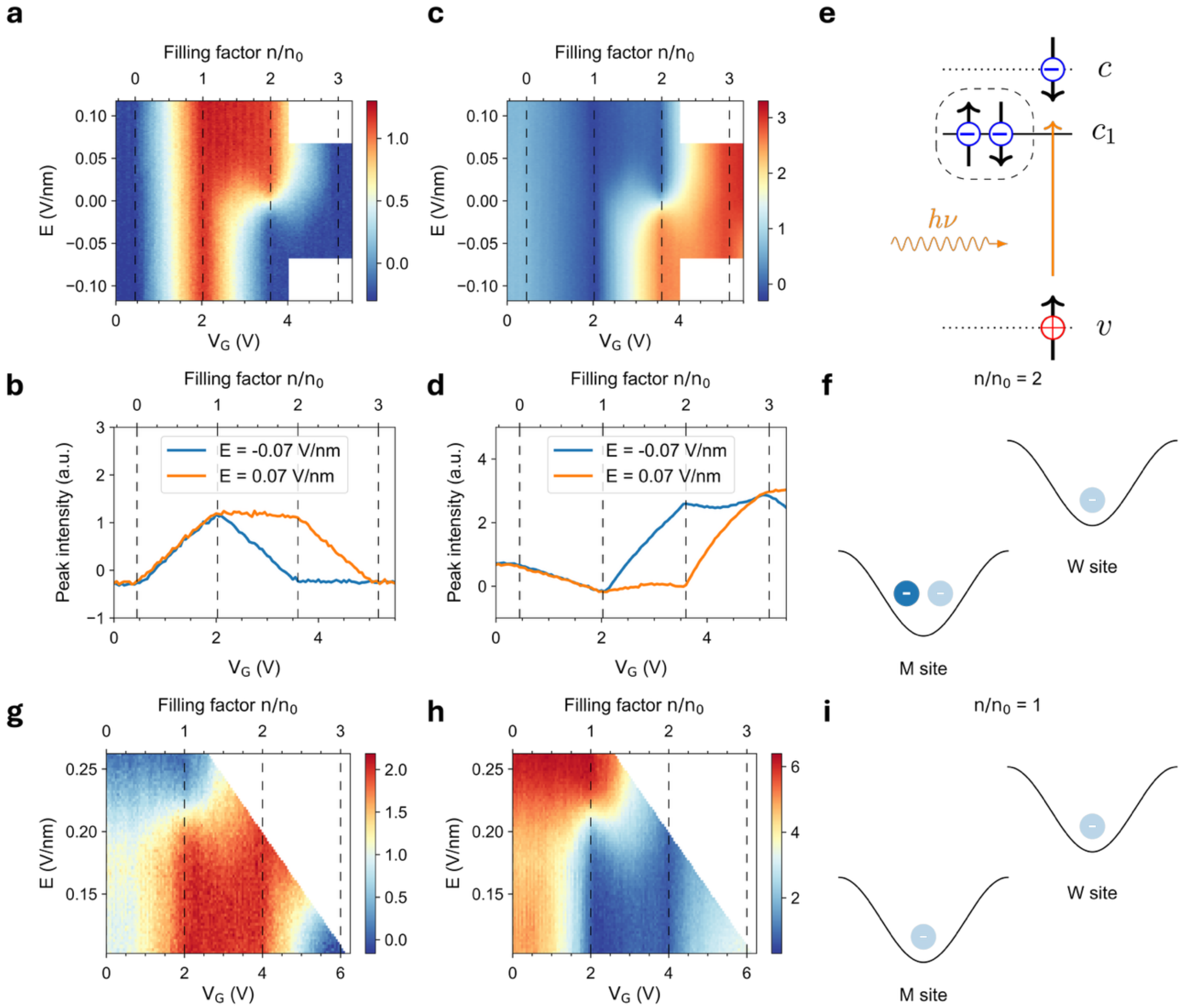

**Fig. 2 | Field-tunable interlayer charge transfer in H-type $MoSe_2/WS_2$ moiré superlattices.** **(a)** Peak intensity of the low-energy moiré trion (LET) state in the doping and vertical electric-field two-dimensional parameter space. **(b)** Two representative horizontal line cuts in (a) with negative (blue) and positive (orange) electric field. **(c)** Peak intensity of the emergent moiré exciton (EX) state in the doping and vertical electric-field two-dimensional parameter space. **(d)** Two representative horizontal line cuts in (c) with negative (blue) and positive (orange) electric field. **(e)** Schematic showing the physical nature of the EX state. **(f)** Schematic showing the charge-transfer process between the M and the W sites in H-type $MoSe_2/WS_2$ moiré superlattices at $n/n_0 = 2$. The transparent electron denotes an electric-field-switchable electron. **(g)** Peak intensity of the LET state in the doping and vertical electric-field two-dimensional parameter space in device D2. **(h)** Peak intensity of the $M_1$ state in the doping and vertical electric-field two-dimensional parameter space in device D2. **(i)** Schematic showing the charge-transfer process between the M and the W sites in H-type $MoSe_2/WS_2$ moiré superlattices at $n/n_0 = 1$. The transparent electron denotes an electric-field-switchable electron.

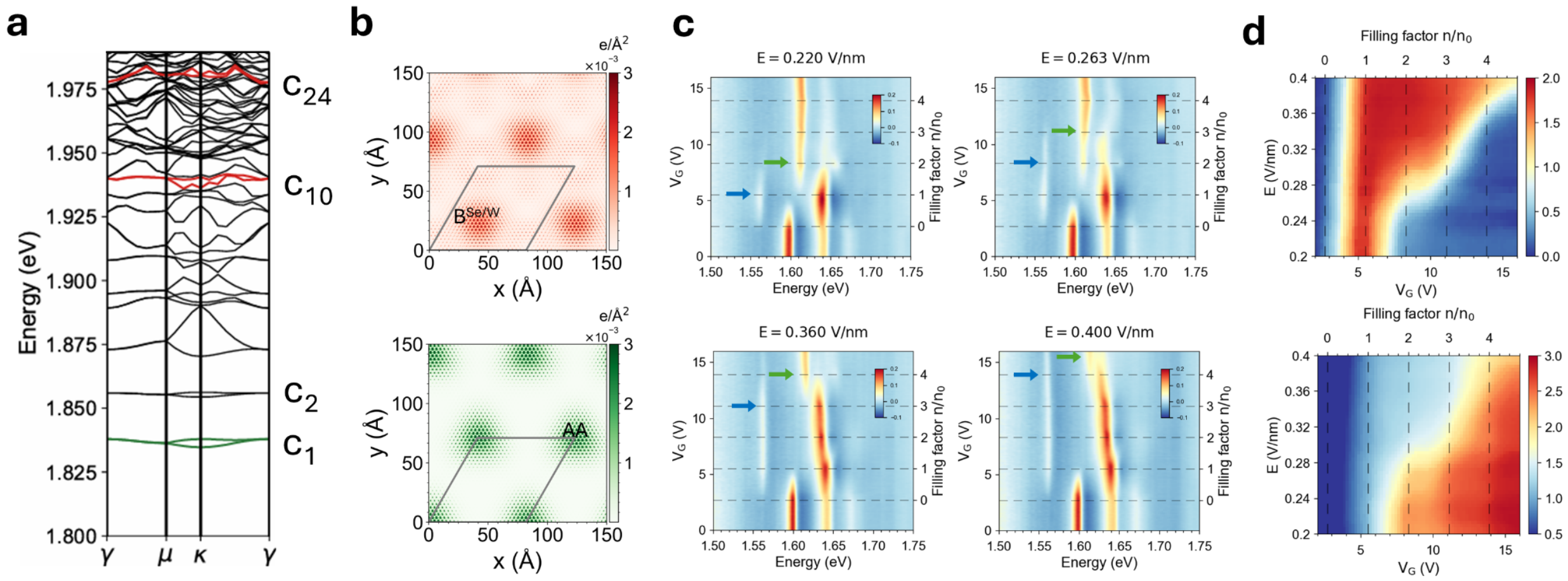

**Fig. 3 | Field-tunable interlayer charge transfer in R-type $MoSe_2/WS_2$ moiré superlattices.** **(a)** Calculated band structure of 0° aligned $MoSe_2/WS_2$ moiré superlattices. $c_1$ and $c_2$ bands are localized in the $MoSe_2$ layer; $c_{10}$ and $c_{24}$ bands are the first two lowest energy conduction states localized in the $WS_2$ layer, derived from the K valley. **(b)** Calculated electron charge density modulation of some of the band states marked in (a). Bottom: first moiré band localized in the $MoSe_2$ layer ($c_1$ band); Top: lowest energy moiré conduction band derived from the K valley that is localized in the $WS_2$ layer ($c_{10}$ band). **(c)** Doping-field RC spectra of the $MoSe_2$ layer with the doping range covering $n/n_0 = 0$ to 4 and the field range covering from 0.22 V/nm to 0.4 V/nm. The blue and green arrows in each spectrum denote the LET and the EX states, respectively. **(d)** Doping-field maps of the $MoSe_2$ moiré excitations. Top: peak intensity of LET; Bottom: peak intensity of EX.

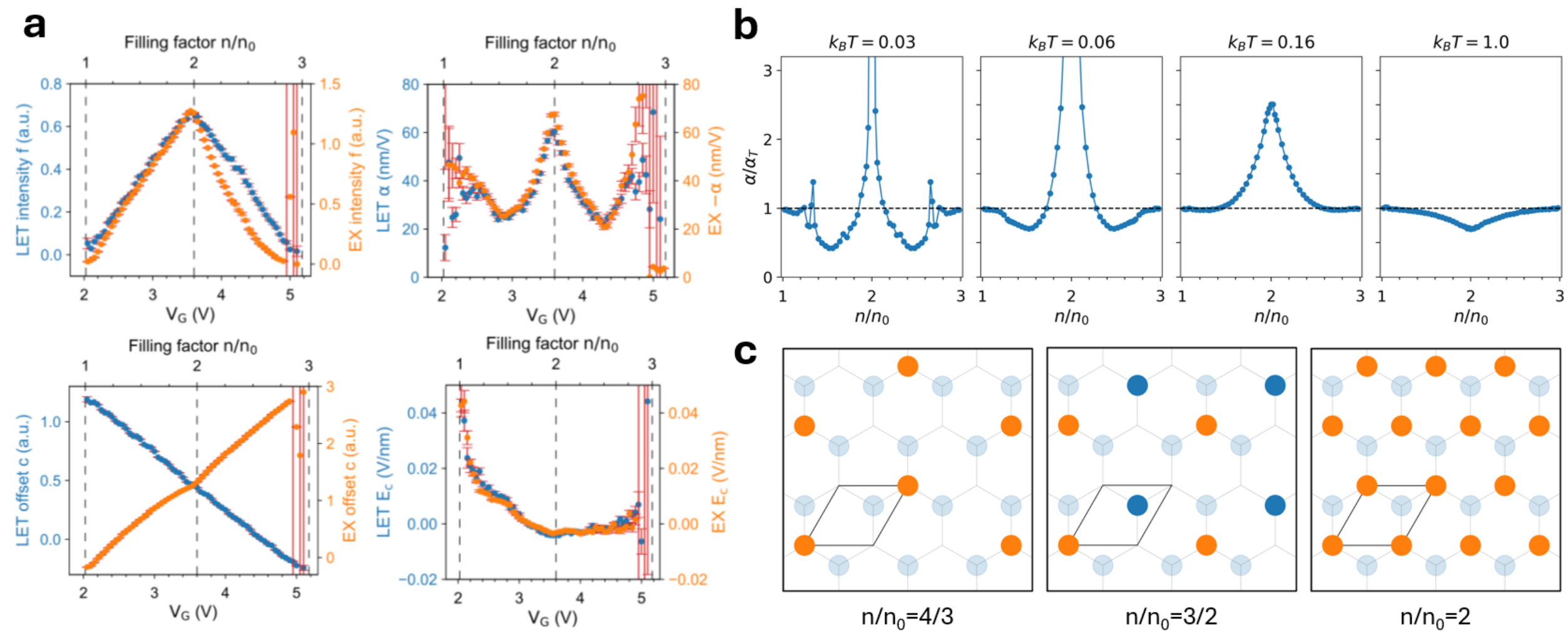

**Fig. 4 | Charge-transfer ordering in H-type $MoSe_2$/$WS_2$ moiré superlattices. (a)** Parameterization of field-dependent intensity of LET (blue) and EX (orange) states from $n/n_0 = 1$ to 3, following the fitting function in the main text. From top left to bottom right: $f$ is the intensity contrast, $c$ is the intensity offset, $\alpha$ is the measured field susceptibility, and $E_c$ is the midpoint critical electric field. **(b)** Simulated doping evolution of $\alpha$ from Monte-Carlo sampling at multiple temperatures. Temperature is normalized against nearest-neighbor M-W Coulomb interaction $V_1$, and $\alpha$ is normalized against non-interacting expectation $\alpha_{\mathrm{T}}$ (see main text). **(c)** Illustration of three observed low-temperature charge orders on the moiré superlattices at $n/n_0 =$ 4/3, 3/2, and 2. The light blue circle represents the $n/n_0 = 1$ background charge at M sites, and solid blue/orange dots represent the electrons localized at the M and the W sites, respectively. The moiré supercell is highlighted to match the M and the W sites with the DFT calculation in Figure 1b.

# Supplementary Information for Tunable Interlayer Charge-transfer States in $MoSe_2/WS_2$ Moiré Superlattices

Zheyu Lu[†,*], Jiahui Nie[†], Tianle Wang[†], Rwik Dutta[†], Ruishi Qi, Jingxu Xie, Can Uzundal, Jianghan Xiao, Ziyu Wang, Yibo Feng, Kenji Watanabe, Takashi Taniguchi, James R. Chelikowsky, Archana Raja, Steven G. Louie, Mit H. Naik, Michael P. Zaletel, Feng Wang[*]

[†]These authors contributed equally.
[*]Correspondence to: zheyulu@berkeley.edu, fengwang76@berkeley.edu

## Second harmonic generation

We use second harmonic generation (SHG) to determine the relative alignment of the TMD flakes. The measurements were performed using an amplified femtosecond laser system based on a commercially available Yb:KGW source (Light Conversion, Carbide), which pumps an optical parametric amplifier (OPA, Light Conversion, Orpheus Twins). The idler output of the OPA, tuned to 1060 nm, was focused onto the sample using a 50× microscope objective. The back-reflected SHG signal was collected and imaged onto a cooled camera (Princeton Instruments, PIXIS-100) using a short focal length lens ($f$ = 2.5 cm). The SHG anisotropy was measured by inserting a polarizer and a half-wave plate into the beam path and recording the SHG intensity as a function of the half-wave plate angle.

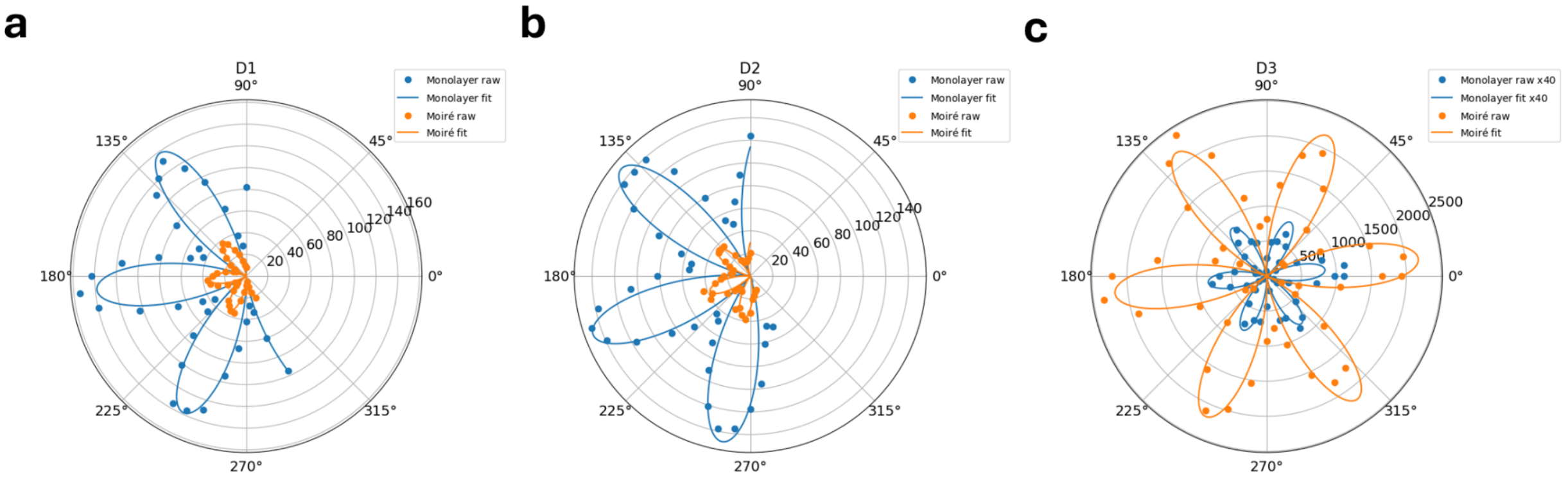

**Supplementary Figure 1 | Second harmonic generation (SHG).** Polarization-resolved SHG data for the three devices described in this study. In the SHG data, the blue (orange) dots correspond to the integrated SHG signal at different polarization angles $\theta$ for the monolayer (moiré) regions, and the solid lines are the corresponding fits to a $\cos^2(3\theta)$ function. **(a)** Device D1: 60°-aligned (H-type) $MoSe_2/WS_2$ heterostructure described in the main text. **(b)** Device D2: 60°-aligned (H-type) $MoSe_2/WS_2$ heterostructure described in the main text. **(c)** Device D3: 0°-aligned (R-type) $MoSe_2/WS_2$ heterostructure described in the main text.

## High-energy moiré trion (HET) state

We propose that the HET state is a trion-like state stabilized by moiré localization, derived from the repulsive polaron state of monolayer $MoSe_2$. The monolayer repulsive polaron is known to be a collective bound state from the exciton dressed by the Fermi sea of free electrons, which is gradually blue shifted by the increase in electron density due to Pauli blocking. However, in the moiré heterobilayer, the strong moiré reconstruction of the single-particle bands leads to a great modification of this collective bound state: the repulsive polaron state can be strongly quantized to integer fillings due to the strong interaction with highly localized electrons in each moiré supercell. Specifically, at $n/n_0 = 1$, the $c_1$ band is half-filled, leading to the Mott-insulating ground state. Each electron can bind with an $M_1$ exciton, forming an attractive polaron state (LET), leaving the other orthogonal repulsive polaron state to the high-energy side. In the simplest single-particle picture, the HET could be viewed as a bound state composed of a higher-energy exciton and a doped electron at the M site ($c_1$ band).

Despite its complicated physical origin, the HET state demonstrates a synchronized emergence/annihilation with the LET state and features identical electric field dependence, as shown in Supplementary Figure 2. Both peaks are associated with the doped electron localized at the M site. As a result, these two emergent excitations form a pair of optical probes capable of detecting a single, localized doped electron in the $MoSe_2$ layer.

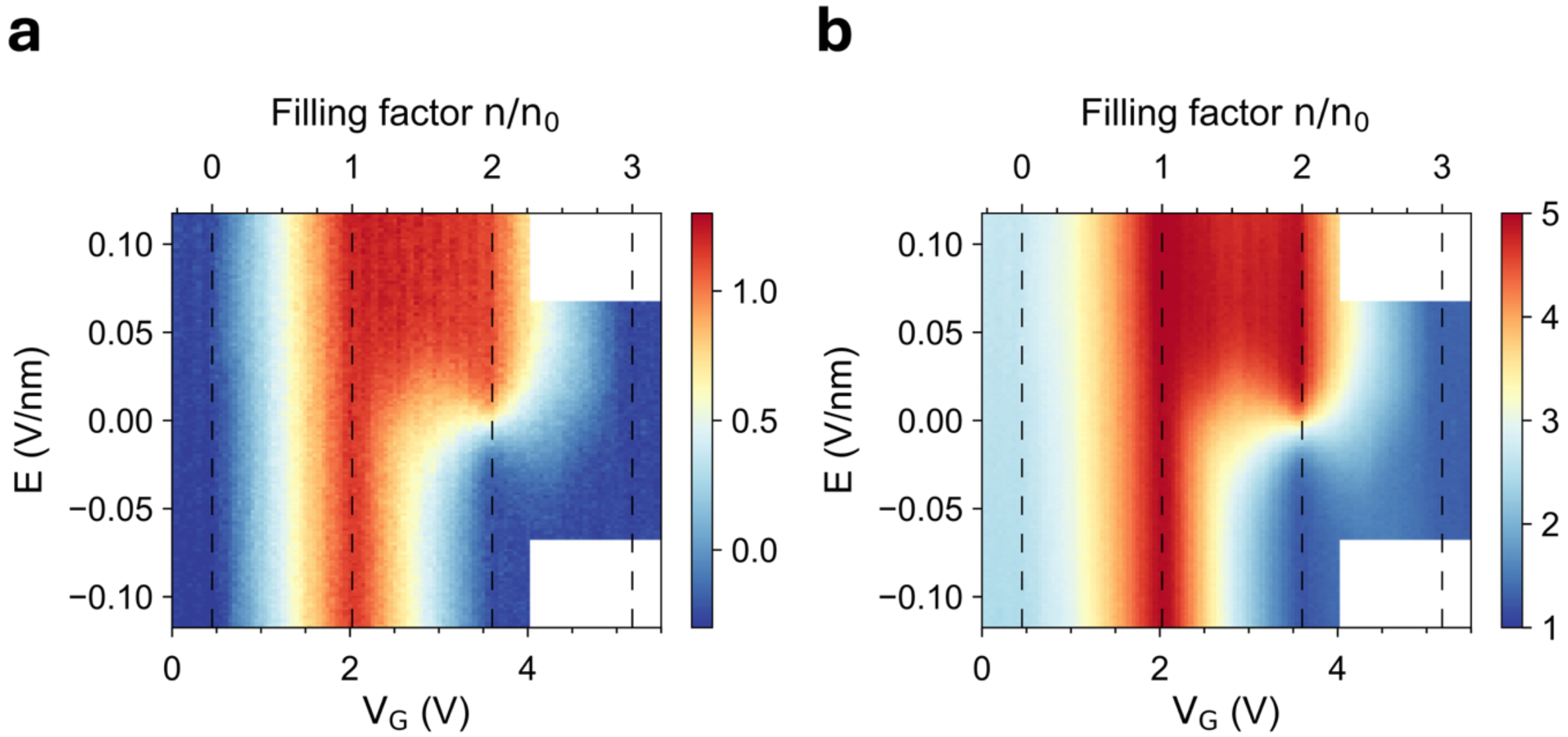

**Supplementary Figure 2 | LET vs HET. (a)** Peak intensity of the low-energy moiré trion (LET) state in the doping and vertical electric-field two-dimensional parameter space. **(b)** Peak intensity of the high-energy moiré trion (HET) state in the doping and vertical electric-field two-dimensional parameter space.

## Additional electron-localization switching behaviors in H-type moiré superlattices

Using the emergent exciton as an optical probe, we extend the study of electron localization to the regime of $n/n_0 = 3$ to 4, where the moiré trion signal is no longer present. The left panel in Supplementary Figure 3a shows the two-dimensional intensity plot of the emergent exciton, which continues to exhibit switching behavior. Supplementary Figure 3b shows the doping-dependent RC spectrum under negative (left) and positive (right) electric fields. Notably, a new lower-energy peak emerges, exhibiting a well-defined magnetic circular dichroism (MCD) signal, as shown in Supplementary Figure 3c. The two-dimensional intensity plot of this new peak is displayed in the right panel of Supplementary Figure 3a. It reveals a complementary pattern to that of the emergent exciton (EX) state. We therefore attribute this new feature to the trion state associated with the EX state, here referred to as the emergent trion (EXT) state, which becomes active when three electrons occupy the M site.

Supplementary Figure 4a shows the two-dimensional map of the EX intensity in device D2, which has a larger electric field range. The blue arrow denotes the electron-localization switching we observed in D1 at $3 < n/n_0 < 4$ (Supplementary Figure 3a). The red arrow, however, denotes a new electron-localization switching at $2 < n/n_0 < 3$ under a large negative electric field. With a large negative electric field, the third doped electron prefers to stay in the $MoSe_2$ layer and fills the $c_2$ band, which is also localized at the M site. As a result, the EX intensity decreases. It is worth noting that although EX intensity decreases under both large positive and negative electric fields at $2 < n/n_0 < 3$, the underlying electron configuration is totally different for the two switching behaviors.

In summary, we have the following additional electron-localization switching behaviors:

(1) $2 < n/n_0 < 3$: By further increasing the electric field in the negative direction, we can put three electrons at the M site, leading to the potential formation of a quantum many-body complex, such as the Wigner molecule state.

(2) $3 < n/n_0 < 4$: At $n/n_0 = 3$ with a moderate electric field, the energetically stable configuration is two electrons forming a spin-singlet at the M site, and one electron occupying the W site. The additional doped electron can be switched between the W and the M sites by occupying the second moiré conduction band ($c_2$ band), which is also localized at the M site.

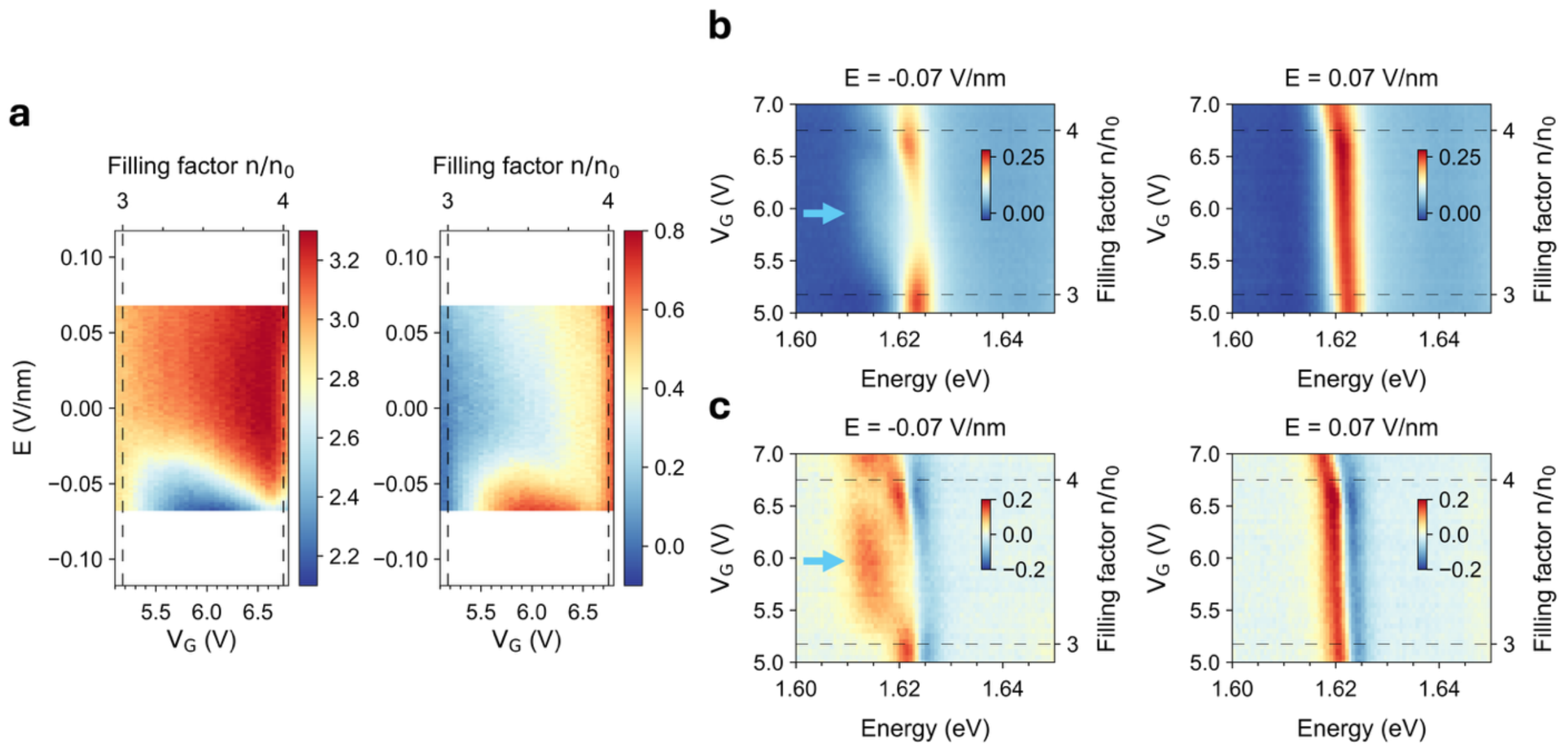

**Supplementary Figure 3 | Electron-localization switching at $3 < n/n_0 < 4$ in H-type $MoSe_2/WS_2$ moiré superlattices. (a)** Left: Peak intensity of emergent moiré exciton (EX) in the doping and vertical electric field two-dimensional parameter space at $3 < n/n_0 < 4$. Right: Peak intensity of emergent trion (EXT) in the doping and vertical electric field two-dimensional parameter space at $3 < n/n_0 < 4$. **(b)** Experimental reflection contrast spectrum of $MoSe_2/WS_2$ moiré superlattices with electron doping ranging from $n/n_0 = 3$ to 4 under a negative (left) and positive (right) electric field, respectively (zoom-in of Figures 1f and 1g in main text). The blue arrow indicates the EXT state. **(c)** The magnetic circular dichroism (MCD) spectrum corresponding to (b).

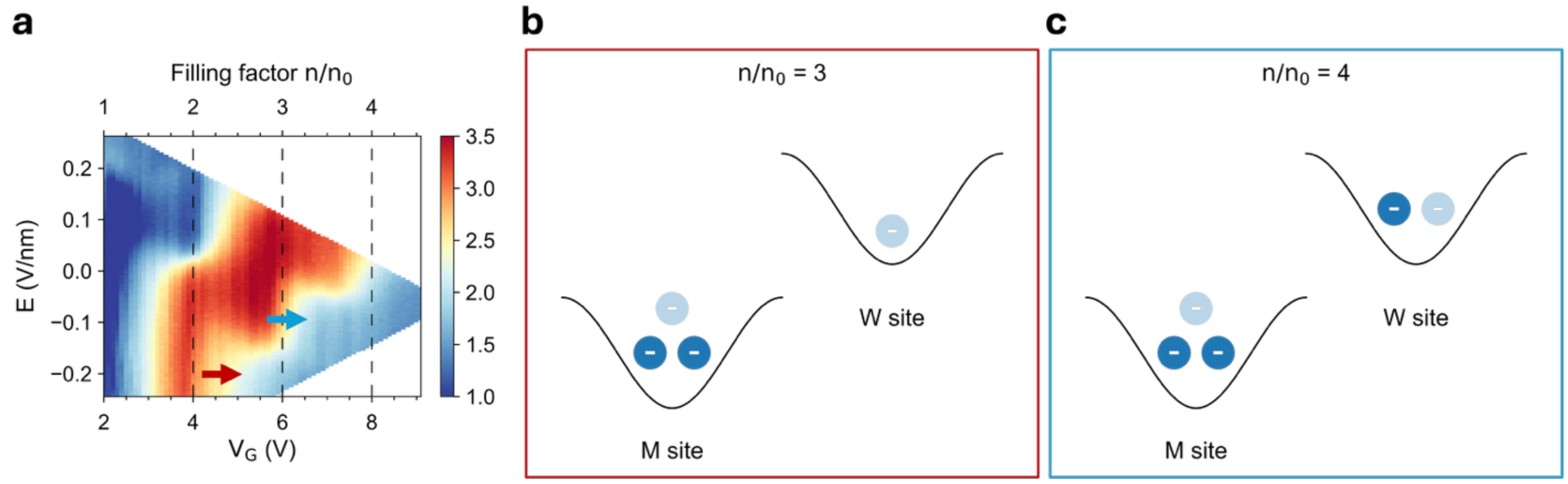

**Supplementary Figure 4 | Additional electron-localization switching behaviors in H-type $MoSe_2/WS_2$ moiré superlattices. (a)** Peak intensity of emergent moiré exciton (EX) in the doping and vertical electric field two-dimensional parameter space in device D2. The red arrow denotes the additional electron-localization switching behaviors at filling $2 < n/n_0 < 3$ under a large negative electric field, in which case the original three-electron configuration under a moderate electric field is no longer stable. Blue arrow denotes the electron-localization switching behaviors at filling $3 < n/n_0 < 4$, consistent with Supplementary Figure 3a. **(b)** Schematic showing the charge-transfer process indicated by the red arrow in (a). The transparent electron denotes an electric-field-switchable electron. **(c)** Schematic showing the charge-transfer process indicated by the blue arrow in (a). The transparent electron denotes an electric-field-switchable electron.

## Electron localization of $n/n_0 = 3$ charge-transfer state in R-type $MoSe_2/WS_2$ moiré superlattices

Here we analyze the possible electron localization for the $n/n_0 = 3$ charge-transfer process in R-type $MoSe_2/WS_2$ moiré superlattices introduced in the main text. The $n/n_0 = 3$ transition appears at $E_{c,3} \approx 0.28$ V/nm. As we show in the main text, at the field below $E_{c,3}$, the $MoSe_2$ side shows a strong EX resonance at $n/n_0 = 3$, while when the field is above $E_{c,3}$, the EX state switches to the LET state. Here, we also include the RC spectrum on the $WS_2$ side. In Supplementary Figure 5a, the RC spectrum on the $WS_2$ side shows a well-defined exciton resonance at around 2.02 eV (pointed by the red arrow) and a trion resonance at a lower energy around 2 eV (pointed by the blue arrow). Supplementary Figure 5b further shows the intensity map of the $WS_2$ trion across the same doping and field range as in the main text (Figure 3d). The $WS_2$ trion persists for electric fields either below or above $E_{c,3}$, indicating that there is always one electron in the $WS_2$ layer across the applied electric field range at $n/n_0 = 3$.

Therefore, the $n/n_0 = 3$ charge-transfer process occurs with EX switching to LET on the $MoSe_2$ side, while the $WS_2$ trion remains unchanged. This scenario gives rise to two possible charge-transfer pathways, as illustrated in Supplementary Figures 5c and 5d. Specifically, one of the two electrons occupying the M site can either transfer to the singly occupied W site or to the unoccupied $B^{Mo/S}$ site (denoted as the W' site).

The first transfer pathway requires a critical electric field $E^{a}_{c,3} \sim (\Delta_0 - U + U_W)/(ed)$, while the second requires $E^{b}_{c,3} \sim (\Delta_1 - U + J_W)/(ed)$. The difference between these critical electric fields is given by $\Delta E_{c,3} = E^{b}_{c,3} - E^{a}_{c,3} \sim (\Delta_1 - \Delta_0) - (U_W - J_W)$, where $\Delta_0$ ($\Delta_1$) denotes the potential energy difference between the $B^{Se/W}$ ($B^{Mo/S}$) stacking, $c_{10}$ ($c_{24}$) band, in the $WS_2$ layer and the AA stacking, $c_1$ band, in the $MoSe_2$ layer. Here, $U_W$ represents the on-site Coulomb repulsion at the $B^{Se/W}$ stacking site, and $J_W$ denotes the Coulomb interaction between $B^{Se/W}$ and $B^{Mo/S}$ sites in the $WS_2$ layer.

From DFT calculations, the $c_{24}$ band is approximately 45 meV higher than the $c_{10}$ band, implying $\Delta_1 - \Delta_0 \sim 45$ meV. The critical field required for the $n/n_0 = 2$ charge-transfer process is approximately $\Delta_0 - U_M$. If the second pathway were energetically favored, the additional field required to reach the $n/n_0 = 3$ state would be $(\Delta_1 - \Delta_0) + J_W > 45$ meV. However, experimentally, the increase in electric field from the $n/n_0 = 2$ to the $n/n_0 = 3$ charge-transfer transition is small (less than 0.1 V/nm), corresponding to an energy scale of roughly 25 meV (we use $\epsilon_{hBN}\epsilon_0 E_{hBN} = \epsilon_{TMD}\epsilon_0 E_{TMD}$, $E_{hBN} \approx (V_T - V_B)/(2d)$, and $\Delta = eE_{TMD}d_{TMD}$, assuming a heterobilayer vertical distance of $d_{TMD} \approx 0.5$ nm and effective out-of-plane dielectric constants $\epsilon_{hBN} = 3.5$ and $\epsilon_{TMD} = 7$). Since this value is significantly smaller than the estimated lower bound, we conclude that the second transfer pathway is energetically unfavorable.

Alternatively, if the first transfer pathway is favored, the required field energies follow the hierarchy: the $n/n_0 = 1$ transition occurs at approximately $\Delta_0$, the $n/n_0 = 2$ transition at $\Delta_0 - U_M$, and the $n/n_0 = 3$ transition lies in between, i.e., $\Delta_0 - U < \Delta_0 - U_M + U_W < \Delta_0$. This implies $U_W < U_M$. Given $\Delta_0 \sim 100$ meV and that the critical field for the $n/n_0 = 2$ transition is about 0.26 V/nm (corresponding to an electron energy of ~ 65 meV), we estimate $U_M$ ~ 35 meV, which leads to $U_W < 35$ meV. Consequently, $U_W - J_W$ must be much smaller than 45 meV (likely in the sub-meV range), yielding $\Delta E_c = E_c^b - E_c^a > 0$, consistent with the first pathway being energetically favored.

We therefore conclude that, for the $n/n_0 = 3$ charge-transfer state, the most likely electron configuration corresponds to the case where one of the two electrons initially occupying the M site transfers to the singly occupied W site. This results in a singly occupied M site and a doubly occupied W site (Supplementary Figure 5c).

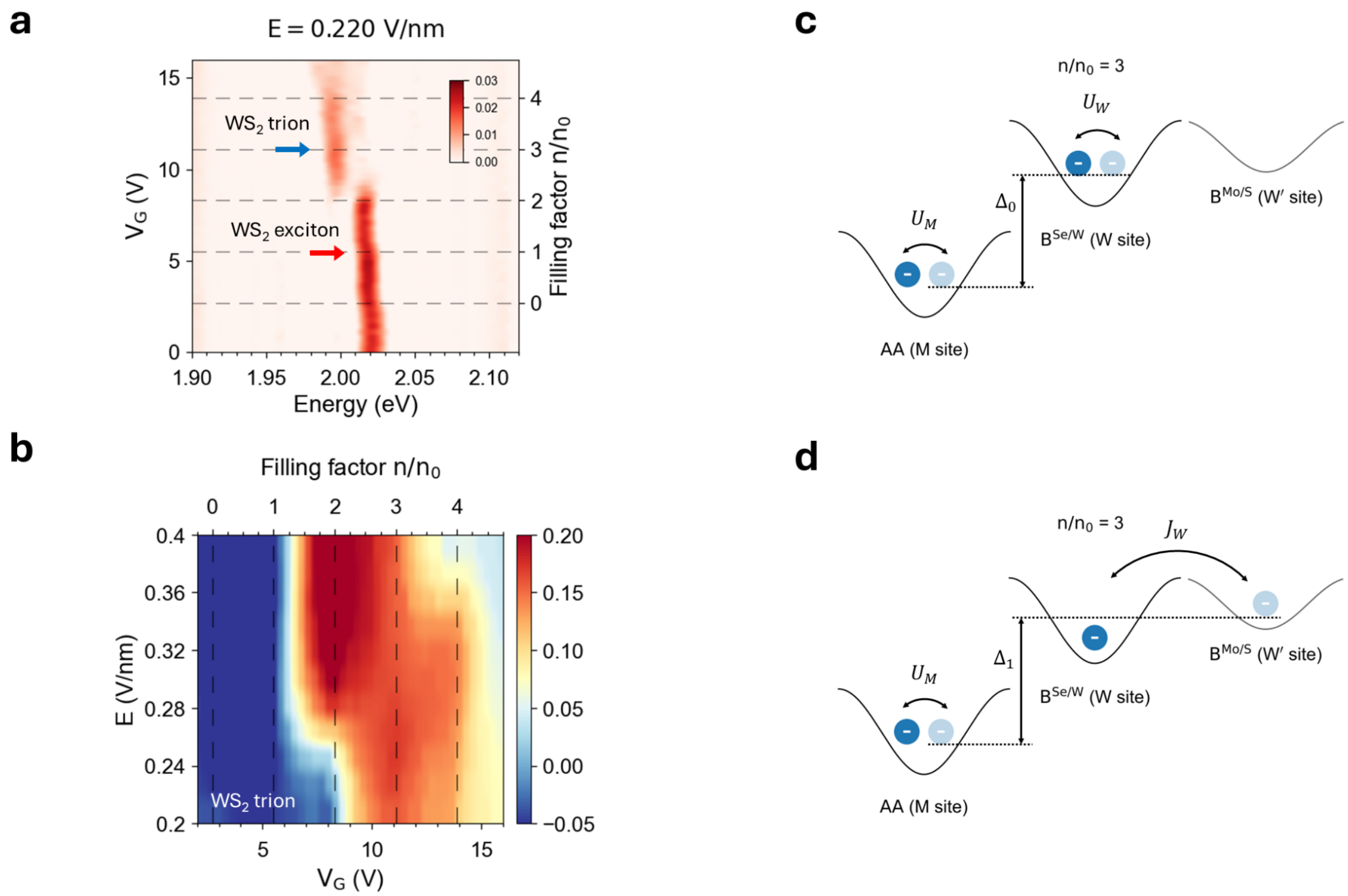

**Supplementary Figure 5 | Schematics of two electron-localization configuration candidates for the $n/n_0 = 3$ charge-transfer state in R-type $MoSe_2/WS_2$ moiré superlattices. (a)** RC spectrum of the $WS_2$ layer at $E = 0.22$ V/nm. The blue arrow denotes the moiré trion state of $WS_2$ at $n/n_0 = 3$. **(b)** Doping-field 2D map of the peak intensity of $WS_2$ trion state. **(c)** Electrons are switched between the M and the W sites. **(d)** Electrons are switched between the M and the W' site.

## Electric field polarization curve in the noninteracting model

In a two-state model without interactions, the LET intensity is given by

$$\begin{aligned} I &\propto N - N_1 \\ &= N - N_\mathrm{e} \cdot \frac{\exp(-\beta E_1)}{\exp(-\beta E_1) + \exp(-\beta E_2)} \\ &= N - N_\mathrm{e}/2 + (N_\mathrm{e}/2) \cdot \tanh\left[\beta(ehE - \Delta)/2\right] \\ &\propto (1 - \delta/2) + (\delta/2) \cdot \tanh\left[\beta(ehE - \Delta)/2\right] \\ &= f \cdot \tanh\left[\alpha(E - E_\mathrm{c})\right] + c \end{aligned}$$

at filling $1 < n/n_0 < 2$, where $N$ is the total number of moiré supercells, $N_1$ is the number of moiré supercells with two electrons at the M site, $N_e$ is the number of additionally doped electrons from $n/n_0 = 1$, $\delta = N_\mathrm{e}/N = \mathrm{n/n_0} - 1$ represents the doping level away from $n/n_0 = 1$, and $E_{1/2} = E_0 \pm (eEh - \Delta)/2$ denote energies of the M and the W sites at electric field $E$, respectively. $E_0$ is the mid-point energy used in the intermediate calculations and does not show up in the final expression. Supplementary Figure 6 shows examples of electric field polarization curves for LET and EX and the corresponding fitting results using the above parametric formula.

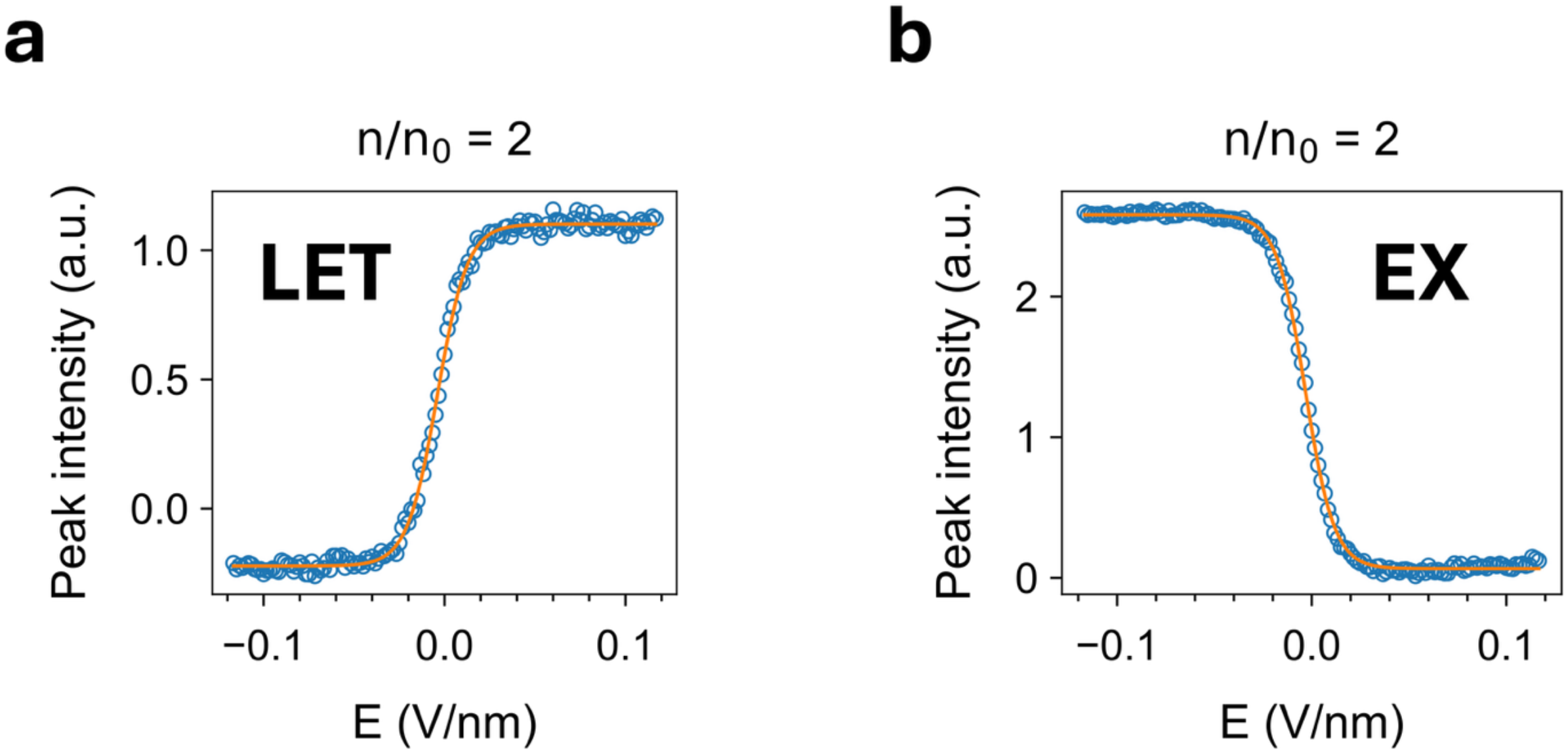

**Supplementary Figure 6 | Electric field polarization curve in H-type $MoSe_2/WS_2$ moiré superlattices. (a)** LET intensity as a function of electric field and its corresponding fitting at $n/n_0 = 2$. **(b)** EX intensity as a function of electric field and its corresponding fitting at $n/n_0 = 2$.

## Theoretical calculations

**Structural Reconstruction:** Aligned bilayer superlattices of $MoSe_2/WS_2$ were constructed at 0° and 60° with the TWISTER code[1]. Experimental lattice constants of 3.28 Å for $MoSe_2$ and 3.15 Å for $WS_2$ were used. Classical force fields parametrized against van der Waals-corrected density functional theory (DFT) data were employed to relax the moiré superlattices. The Stillinger-Weber[2] and Kolmogorov–Crespi[3,4] potentials were used to describe intralayer and interlayer interactions, respectively, as implemented within the LAMMPS package[5].

**Density functional theory:** Density functional theory (DFT) calculations[6] of the reconstructed $MoSe_2/WS_2$ moiré superlattice were performed with the SIESTA code[7]. The exchange-correlation effects were treated using generalized gradient approximation following the Perdew-Burke-Ernzerhof (PBE) implementation[8]. Fully relativistic norm-conserving pseudopotentials[9,10] were employed to describe the interactions between valence electrons and ions. The self-consistent ground-state charge density was obtained by $\gamma$ point sampling of the moiré Brillouin zone. The charge density was then used to construct the DFT Hamiltonian at arbitrary $k$-points across the moiré Brillouin zone, yielding the band structure and wavefunctions needed for subsequent excited-state calculations. Spin-orbit coupling was explicitly included in all calculations.

Pristine unit-cell DFT calculations for the $MoSe_2$ layer were performed using the Quantum Espresso package[11] using identical pseudopotentials and exchange-correlation functional to the SIESTA calculations. The single-particle Kohn-Sham states were expanded in plane waves with a kinetic-energy cutoff of 40 Ry. The Brillouin zone was sampled with a 12×12×1 $k$-point grid to obtain the ground state charge density in an SCF cycle.

**GW-BSE:** DFT reasonably describes the band dispersions but underestimates the bandgap. To correct the bandgap, we evaluated the quasiparticle energies within the GW self-energy approximation[12] for monolayer $MoSe_2$, as implemented in the BerkeleyGW package[13]. The DFT conduction bands of the reconstructed moiré superlattice were rigidly shifted using the resulting GW correction to the band gap at the K point. The static dielectric function $\epsilon$ and screened Coulomb interaction $W$ were evaluated within the random phase approximation, employing a plane-wave energy cutoff of 35 Ry and ~ 5000 unoccupied states. Extension of the dielectric function to finite frequencies was performed using the Hybertsen-Louie generalized plasmon-pole model[12]. The Coulomb potential was truncated in the out-of-plane direction to exclude the spurious effects of periodic images[14]. Convergence of the $q$-point sampling was enhanced through the nonuniform neck subsampling (NNS) technique[15].

Electron-hole correlations were explicitly treated by the electron-hole interaction kernel in the Bethe–Salpeter equation (BSE)[16]. Brute force computation of the moiré electron-hole kernel is computationally intractable due to the large dimensionality of the moiré electronic wavefunctions. Therefore, the pristine unit-cell matrix projection (PUMP) method[17,18] was employed, which is described in detail below. PUMP is used to construct the moiré superlattice kernel by expressing it as a coherent linear combination of multiple pristine unit-cell kernel matrix elements. The BSE Hamiltonian for the reconstructed $MoSe_2$ superlattice was constructed using 24 valence and 24 conduction bands, with 3×3×1 sampling of the moiré Brillouin zone. These moiré wavefunctions were expanded in terms of 48 valence and conduction states of the pristine superlattice.

Screening was obtained for an AA-stacked $MoSe_2/WS_2$ heterobilayer in the primitive unit-cell to account for the effects of both the layers. The interlayer spacing was chosen to be 6.6 Å which is the average interlayer spacing found in the moiré superlattice. The static dielectric matrix, to compute the unit-cell kernel matrix elements was expanded in plane waves up to 4 Ry and 364 unoccupied states. These pristine unit-cell kernel matrix elements were used to construct the moiré BSE Hamiltonian (more details below) which was diagonalized to obtain the exciton energy levels $\Omega_S$ and the exciton envelope function $A^{S}_{cvk}$. We then compute the imaginary part of the dielectric function, $\epsilon_2(\omega)$ and the absorption $A(\omega)$ (Supplementary Figure 9). In the weakly absorbing limit, the absorption is expressed as $A(\omega) = \frac{\epsilon_2(\omega)\omega d}{c}$ where $d$ is the sample thickness equal to the average bilayer thickness of 9.8 Å.

**Pristine unit-cell matrix projection method:** To obtain the moiré excitonic states, we solve the BSE:

$$\left(E^{QP}_{\mathbf{ck}} - E^{QP}_{\mathbf{vk}}\right)A^{S}_{\mathbf{cvk}} + \sum_{v'c'k'} \langle \psi^{SL}_{\mathbf{vk}} \psi^{SL}_{\mathbf{ck}} | K^{eh} | \psi^{SL}_{v'\mathbf{k}'} \psi^{SL}_{c'\mathbf{k}'} \rangle A^{S}_{c'v'\mathbf{k}'} = \Omega_S A^{S}_{\mathbf{cvk}}, \qquad (1)$$

where $E^{QP}_{\mathbf{nk}}$ denotes the quasiparticle eigenvalues, $\psi^{SL}_{\mathbf{nk}}$ are the superlattice wavefunctions, $A^{S}_{\mathbf{cvk}}$ and $\Omega_S$ are the exciton eigenfunctions and eigenvalues, respectively. The indices v, c, **k**, correspond to the valence band index, conduction band index, and crystal momentum, respectively. The electron-hole interaction kernel $K^{eh}$ comprises an attractive direct term involving the screened Coulomb interaction and a repulsive exchange term. Evaluation of the kernel matrix elements scales as $N^5$ with system size, making this a computational bottleneck for large moiré superlattices. The pristine unit-cell matrix projection (PUMP) method reformulates this problem by expressing moiré kernel as a linear combination of pristine unit-cell kernel matrix elements, yielding a substantial reduction in computational cost.

In the PUMP method, the superlattice electronic wavefunctions are expanded in the basis of primitive unit-cell electronic states. Each valence ($|\psi^{\mathrm{SL}}_{\mathbf{vk}}\rangle$) and conduction ($|\psi^{\mathrm{SL}}_{\mathbf{ck}}\rangle$) state in the reconstructed $MoSe_2$ superlattice is expressed as a linear combination of many states of the pristine $MoSe_2$ superlattice: $|\psi^{\mathrm{SL}}_{\mathbf{vk}}\rangle = \sum_i a_i^{\mathbf{vk}} |\Phi^{\mathrm{val}}_{i\mathbf{k}}\rangle$, $|\psi^{\mathrm{SL}}_{\mathbf{ck}}\rangle = \sum_i a_i^{\mathbf{ck}} |\Phi^{\mathrm{cond}}_{i\mathbf{k}}\rangle$, where $\Phi^{\mathrm{val}}_{i\mathbf{k}}$ and $\Phi^{\mathrm{cond}}_{i\mathbf{k}}$ refer to the pristine superlattice wavefunctions of the valence and conduction states, respectively. The pristine superlattice states are in turn related to states in the primitive unit-cell BZ via band folding. We approximate each electron-hole interaction moiré kernel matrix element, $\mathrm{K}^{\mathrm{eh}}$, of the BSE as a coherent sum over many pristine unit-cell kernel matrix elements,

$$\left\langle \psi^{\mathrm{SL}}_{\mathbf{vk}} \psi^{\mathrm{SL}}_{\mathbf{ck}} \middle| \mathrm{K}^{\mathrm{eh}} \middle| \psi^{\mathrm{SL}}_{\mathbf{v'k'}} \psi^{\mathrm{SL}}_{\mathbf{c'k'}} \right\rangle$$

$$\approx \sum_{ijpq} a_i^{\mathbf{vk}*} a_j^{\mathbf{ck}*} a_p^{\mathbf{vk}} a_q^{\mathbf{ck}} \left\langle \Phi^{\mathrm{val}}_{i\mathbf{k}_\mathrm{m}} \Phi^{\mathrm{cond}}_{j\mathbf{k}_\mathrm{m}} \middle| \mathrm{K}^{\mathrm{eh}} \middle| \Phi^{\mathrm{val}}_{p\mathbf{k}'_\mathrm{m}} \Phi^{\mathrm{cond}}_{q\mathbf{k}'_\mathrm{m}} \right\rangle, \quad (2)$$

$$\approx \sum_{ijpq} a_i^{\mathbf{vk}*} a_j^{\mathbf{ck}*} a_p^{\mathbf{vk}} a_q^{\mathbf{ck}} \left\langle \phi^{\mathrm{val}}_{s\mathbf{k}^1_\mathrm{uc}} \phi^{\mathrm{cond}}_{t\mathbf{k}^2_\mathrm{uc}} \middle| \mathrm{K}^{\mathrm{eh}} \middle| \phi^{\mathrm{val}}_{y\mathbf{k}^3_\mathrm{uc}} \phi^{\mathrm{cond}}_{z\mathbf{k}^4_\mathrm{uc}} \right\rangle, \quad (3)$$

where $i$ and $p$ are pristine valence band indices, $j$ and $q$ are pristine conduction band indices in the superlattice. The pristine states $i\mathbf{k}_\mathrm{m}, j\mathbf{k}_\mathrm{m}, p\mathbf{k}'_\mathrm{m}$ and $q\mathbf{k}'_\mathrm{m}$ in the superlattice BZ are related to $s\mathbf{k}^1_\mathrm{uc}, t\mathbf{k}^2_\mathrm{uc}, y\mathbf{k}^3_\mathrm{uc}$ and $z\mathbf{k}^4_\mathrm{uc}$ in the unit-cell BZ by band folding, respectively. The kernel in eqn. 2 corresponds to the pristine supercell matrix element, while the kernel matrix elements in eqn. 3 refers to the pristine unit-cell matrix element.

**Moiré emergent exciton (EX) state:** The EX peak appears in the optical spectrum when two electrons occupy the M site, corresponding to a ground state in which the $c_1$ band is fully filled. To determine the real-space distribution of this EX state (Supplementary Figure 10), we construct the moiré kernel (eqn. 2) after excluding contributions from the $c_1$ moiré states. In this procedure, we retain the same screened Coulomb interaction $W$, neglecting changes in screening corresponding to the removal of the $c_1$ band.

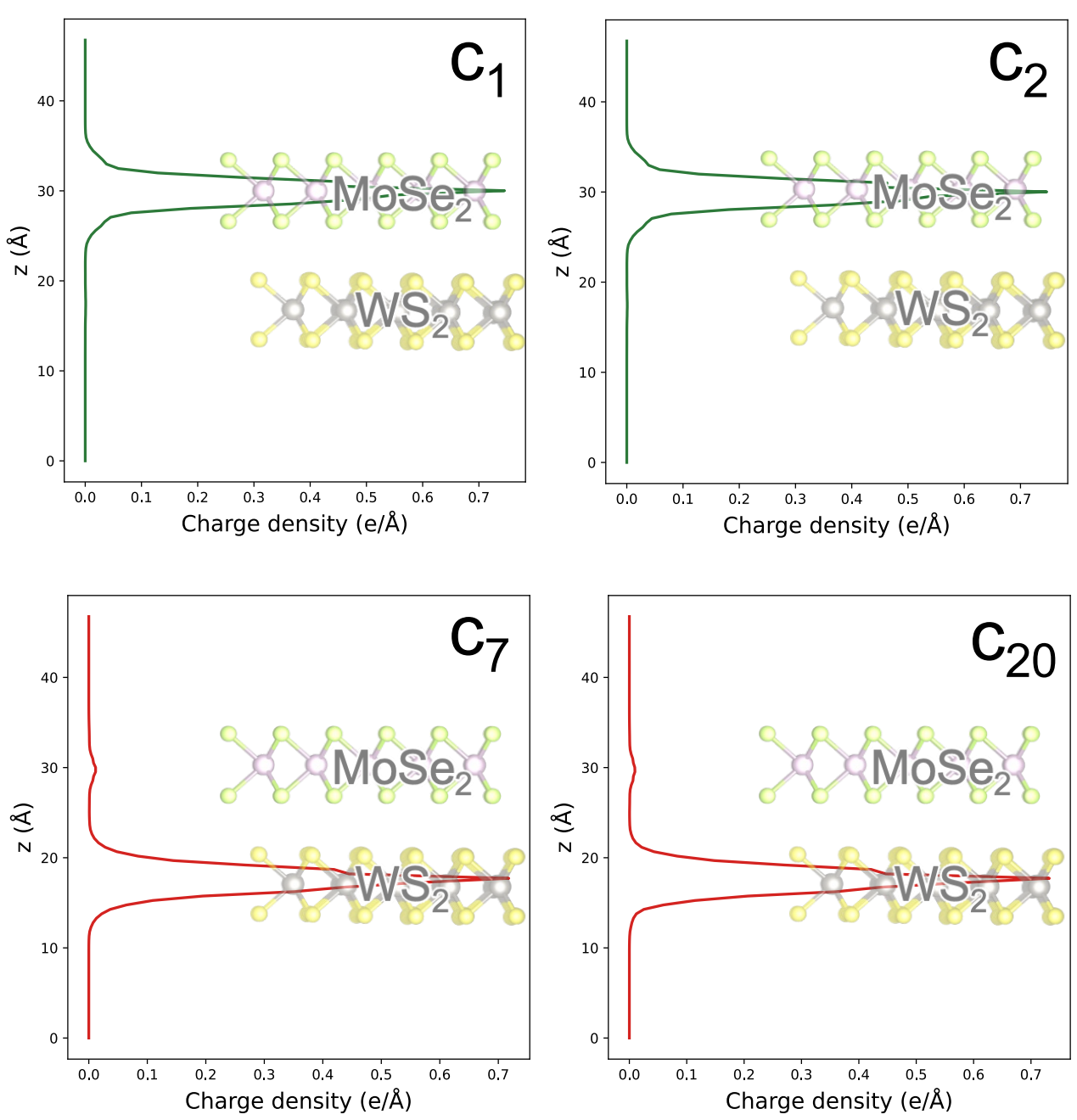

**Supplementary Figure 7 | Out-of-plane electron density of H-type $MoSe_2/WS_2$ moiré superlattices.** $c_1$ and $c_2$ bands are localized in the $MoSe_2$ layer; $c_7$ and $c_{20}$ bands are the first two lowest energy conduction band states derived from the K valley that are localized in the $WS_2$ layer.

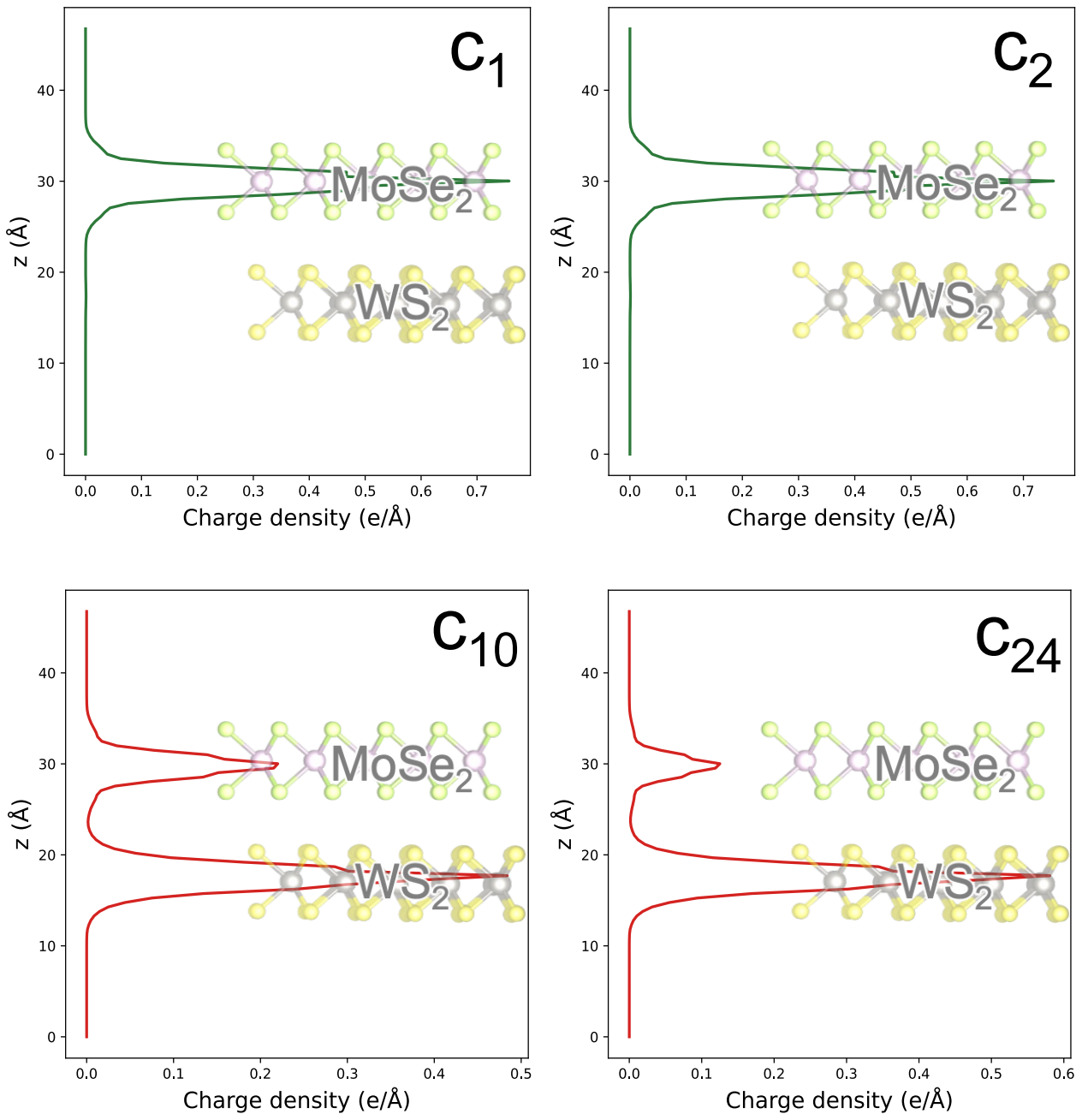

**Supplementary Figure 8 | Out-of-plane electron density of R-type $MoSe_2/WS_2$ moiré superlattices.** $c_1$ and $c_2$ bands are localized in the $MoSe_2$ layer; $c_{10}$ and $c_{24}$ bands are the first two lowest energy conduction band states that are derived from the K valley of the $WS_2$ layer. We observe some mixing of the $c_{10}$ and $c_{24}$ bands with the $MoSe_2$ states in the R-type configuration.

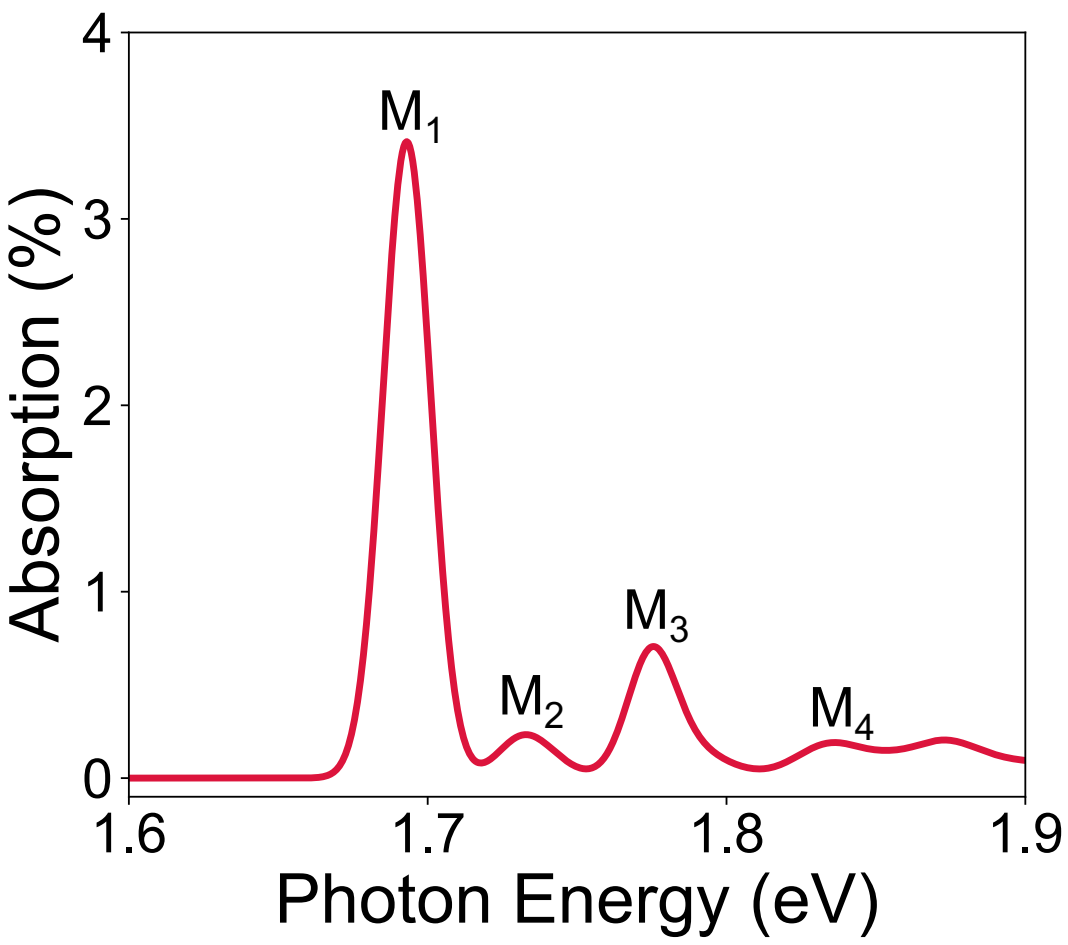

**Supplementary Figure 9 | Calculated absorption spectrum.** Absorption spectrum of H-type $MoSe_2/WS_2$ moiré superlattices, calculated using first-principles GW-BSE calculations with the PUMP method.

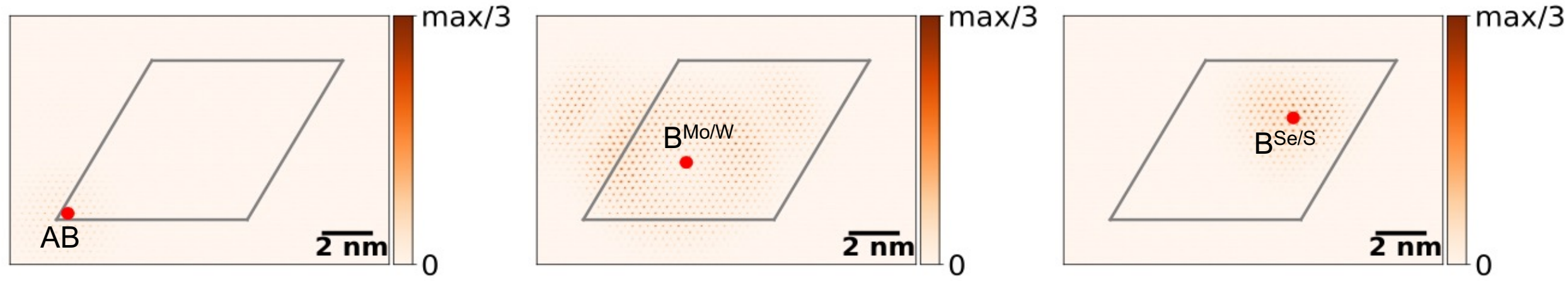

**Supplementary Figure 10 | Real-space map of the EX state in H-type $MoSe_2/WS_2$ moiré superlattices.** The plot shows the electron density of the exciton, $|\chi(\mathbf{r_e}, \mathbf{r_h})|^2$, for fixed hole positions $\mathbf{r_h}$ (red dot) in the moiré superlattice.